\documentclass[12pt,preprint,showpacs,aps,prd,nofootinbib,showkeys,unsortedaddress,floatfix]{revtex4-1}
\pdfoutput=1

\usepackage{makecell}
\usepackage{bm}
\usepackage{amsmath}
\usepackage{graphicx}
\usepackage[usenames,dvipsnames]{color}
\definecolor{darkblue}{RGB}{0,0,196}
\usepackage[colorlinks=true,linkcolor=darkblue,citecolor=darkblue,urlcolor=darkblue]{hyperref}

\newcommand{\beq}{\begin{equation}}
\newcommand{\eeq}{\end{equation}}
\newcommand{\bea}{\begin{eqnarray}}
\newcommand{\eea}{\end{eqnarray}}

\begin{document}

\title{Photon production from a non-equilibrium quark-gluon plasma}

\author{Lusaka Bhattacharya} 
\affiliation{Department of Physics, Kent State University, Kent, OH 44242 United States}

\author{Radoslaw Ryblewski}
\affiliation{The H. Niewodnicza\'nski Institute of Nuclear Physics, Polish Academy of Sciences, 
PL-31342 Krak\'ow, Poland} 

\author{Michael Strickland} 
\affiliation{Department of Physics, Kent State University, Kent, OH 44242 United States}

\begin{abstract}
We calculate leading-order medium photon yields from a quark-gluon plasma using (3+1)-dimensional anisotropic hydrodynamics. Non-equilibrium corrections to the photon rate are taken into account using a self-consistent modification of the particle distribution functions and the corresponding anisotropic hard-loop fermionic self-energies.  We present predictions for the high-energy photon spectrum and photon elliptic flow as a function of transverse momentum, shear viscosity, and initial momentum-space anisotropy.   Our findings indicate that high-energy photon production is sensitive to the assumed level of initial momentum-space anisotropy of the quark-gluon plasma. As a result, it may be possible to experimentally constrain the early-time momentum-space anisotropy of the quark-gluon plasma generated in relativistic heavy-ion collisions using high-energy photon yields.
\end{abstract}


\maketitle 

%%%%%%%%%%%%%%%%%%%%%%%%%%%%%%%%%%%%%%%%%%%%%%%%%%%%%%%%%%%%%%%%%%%%%%%%%%%%%%%%%%%%%%%%%%%%%%%%
\section{Introduction}
%%%%%%%%%%%%%%%%%%%%%%%%%%%%%%%%%%%%%%%%%%%%%%%%%%%%%%%%%%%%%%%%%%%%%%%%%%%%%%%%%%%%%%%%%%%%%%%%

The goal of the ongoing relativistic heavy-ion collision 
experiments at Brookhaven National Laboratory's Relativistic Heavy Ion Collider (RHIC) and at CERN's 
Large Hadron Collider (LHC) is to produce and study the properties of 
the quark-gluon plasma (QGP). It is now well accepted 
that a few microseconds after the Big Bang the entire universe consisted of an extremely hot and dense QGP.  In order to reproduce these conditions terrestrially, 
relativistic heavy-ion collisions are used.  As a result, extremely high 
temperatures and energy densities are created within a very small volume ($\sim 4000 \, {\rm fm}^3$). At these high temperatures, quarks and gluons no longer remain 
confined within nucleons and one instead generates deconfined nuclear matter called a QGP. 
Immediately after the initial nuclear impact, the QGP generated in relativistic heavy-ion collisions 
cools by expansion and, below a certain (pseudo-)critical temperature ($T_c\sim 165$ MeV), the quarks and gluons 
recombine to form hadrons.  After the transition to
hadrons, the system may undergo further expansion and 
cooling before full chemical and kinetic freeze-out takes place.  The resulting particle production and associated radiation from the event are then analyzed in order to infer information about the properties of the QGP.

One of the key outstanding questions in the study of the QGP is the question of the time
scale for the thermalization and isotropization of the matter created in relativistic
heavy-ion collisions.  Theoretical calculations in both the weak-coupling and strong-coupling
limits find that the QGP created immediately after the initial nuclear impact ($\tau \sim 0.2$ fm/c) is highly 
anisotropic in local rest frame (LRF) momentum, however, there are currently no clear experimental
observables that can be used to confirm this expectation and constrain the degree of early-time 
momentum-space anisotropy.  For this purpose, radiation of photons and dileptons are promising
signals since they can be used to probe the initial state of heavy-ion 
collisions. Unlike hadrons, which are emitted from the freeze-out surface 
after undergoing intense re-scatterings, photons emerge from all phases of the expanding fireball:
initial hard scatterings, pre-equilibrium phase, near-equilibrium phase, and hadronic phase. 
Since photons are electromagnetic probes, 
they interact only weakly with the QGP ($\alpha \ll \alpha_s$) and their mean free path is much larger than the 
typical system size ($\sim 10$ fm). As a result, once produced, they do not 
undergo significant interactions with the medium and carry 
largely undistorted information about the circumstances of their production to the 
detector. 

For the most part, in the past calculations of the QGP photon production rate have been 
performed assuming a perfectly thermalized, weakly-coupled QGP when using hydrodynamics
for the background evolution.
Within this framework a complete calculation
of the thermal photon rate at O($e^2{g_s}^2$) has been available for a decade~\cite{Arnold:2001ms} and the next-to-leading-order (NLO) correction O($e^2{g_s}^3$) 
to thermal rate has been computed recently~\cite{Ghiglieri:2013gia}.  At low-temperatures, below the pseudo-critical temperature for the QCD phase 
transition, where dense QCD matter can be
modeled as a hadron resonance gas, effective Lagrangian
approaches have been used, see e.g. Ref.~\cite{Turbide:2003si}.
The success of viscous hydrodynamics applied to heavy-ion collisions~\cite{Heinz:2013th,Gale:2013da} suggests that it might be reasonable to assume that the medium is close to being in local thermal equilibrium. However, nonzero values of the QCD transport coefficients, resulting from nonzero mean free 
paths of the constituents, lead to deviations from local thermal 
equilibrium which increase with the local expansion rate.

For example, in a dissipative QGP, a finite shear
viscosity causes the momentum distribution in the LRF 
to become highly anisotropic at early times, with the distribution falling off more rapidly in
the directions in which the system expands (longitudinal cooling).
In the past, various attempts have been made to determine the
effect of viscous corrections on the photon emission rates in a QGP
~\cite{Dusling:2008xj, Dusling:2009bc, Dion:2011pp}. However, these 
previous works have a key shortcoming: They include only the 
viscous corrections to the local momentum distribution functions 
for the incoming and outgoing particles, but ignore viscous
medium modification of the collision matrix element itself.
For scattering processes in which the inclusion of medium
effects is essential, one must self-consistently include viscous corrections.
For example, when dynamical mass
generation for the medium constituents serves as a regulator
for IR divergences, viscous corrections to the distribution
functions can lead to significant modifications of the screening
mechanism and, therefore, to the collision matrix element itself. This
problem was first addressed using a spheroidal form for the 
LRF momentum-space anisotropy in Ref.~\cite{Schenke:2006fz} with details and followup
studies presented in Refs.~\cite{Schenke:2006yp,Bhattacharya:2008up,Bhattacharya:2008mv,Bhattacharya:2009sb}.

Recently it has been shown how to extend the methods used in Ref.~\cite{Schenke:2006fz} to include the full shear-stress tensor modification of the one-particle distribution function \cite{Shen:2013vja,Shen:2013cca,Shen:2014nfa,Shen:2015qba}.
The calculation was based on photon production including Compton scattering and quark-antiquark 
annihilation at leading order in $\alpha_s$.  Extending the proof presented originally in Refs.~\cite{Schenke:2006fz,Schenke:2006yp}, Shen, Paquet, Heinz, and Gale were able to show that the Kubo-Martin-Schwinger relation holds in the hard-loop (HL) regime for any particle momentum distribution function that is reflection symmetric.\footnote{One must also require that $f_q = f_{\bar{q}}$, which was implicit in their proof.}  Using this, they were able to compute the hard and soft contributions to the rate separately, taking into account the modifications to the (anti-)quark self-energy necessary to make it consistent with a standard Grad-14 $\delta f$ modification to the one-particle distribution function.  

One potential problem with using the standard Grad-14 form for the viscous correction to the particle distribution function is that, in the integrals that determine the photon production rate, one is integrating over all momenta.  At high-momentum, the standard viscous approximation to the particle distribution function is not reliable and, as a result, one finds regions in the integration domain where the viscosity-corrected distribution function is negative.  In this paper, we use instead  a spheroidally-deformed form of the non-equilibrium distribution function that is positive by construction.  We concentrate on photon production from the deconfined phase of the QGP's lifetime and extend prior results  
performed in Refs.~\cite{Schenke:2006yp,Schenke:2008hw,Bhattacharya:2008up} to include a 
more realistic QGP background evolution.  In order to improve upon previous works~\cite{Schenke:2006yp,Schenke:2008hw,Bhattacharya:2008up}, in which a simple analytic model with an adjustable isotropization time was used for the QGP evolution, in this paper we use leading-order (3+1)-dimensional anisotropic hydrodynamics ((3+1)D aHydro)~\cite{Martinez:2010sc,Florkowski:2010cf,Tinti:2014yya,Bazow:2015cha} to describe the non-equilibrium QGP evolution.\footnote{For a recent review of anisotropic hydrodynamics see Ref.~\cite{Strickland:2014pga}.}  
Herein, we use the aHydro
equations obtained from the zeroth and first moments
of the Boltzmann equation with the collisional kernel
treated in the relaxation-time approximation.  The resulting dynamical equations
describe the full (3+1)D spatiotemporal evolution of the transverse
momentum scale $\Lambda$ and spheroidal momentum-space
anisotropy parameter 
$\xi$~\cite{Romatschke:2003ms,Romatschke:2004jh}. 
The (3+1)D
framework allows $\Lambda$, $\xi$, and the associated flow
velocities to depend arbitrarily on the transverse coordinates,
spatial rapidity, and longitudinal proper-time,
however, herein we restrict ourselves to smooth Glauber-like
initial conditions.

We also mention that there there have been recent studies that have used various types of partonic and hadronic kinetic transport codes to address the problem of photon production in heavy-ion collisions \cite{Linnyk:2013hta,Linnyk:2013wma,Linnyk:2015nea,Linnyk:2015tha}.  Such codes automatically take into account the non-equilibrium quark and gluon phase space distributions, but they ignore the effect of non-equilibrium (anisotropic) screening in the problem.  Typically in the kinetic approaches it is assumed that there is a cutoff on $u$- and $t$-channel exchanges which is either fixed or dynamically set by the local Debye mass determined from the parton density.  As shown in Refs.~\cite{Schenke:2006fz,Schenke:2006yp}, in order to properly regulate the infrared divergences in the photon production calculation, it is necessary to revisit the calculation of the of screening and use a quark self-energy which is consistent with the non-equilibrium (locally anisotropic) nature of the QGP.  That being said, it may be that, within the accuracy required for QGP phenomenology, it is sufficient to simply have an isotropic infrared cutoff that does not take into account the non-equilibrium nature of the quark distribution function.  In such a case, the transport codes would provide a quite reasonable calculation of photon production.

The study presented in this paper is similar in spirit to other
studies of photon production using second-order viscous hydrodynamics. 
Our goal is not to produce the most complete calculation of photon production from all possible sources and including fluctuating initial conditions etc., but to instead study the effect of self-consistently including the non-equilibrium modifications of the quark distribution function and to systematically investigate the dependence of the resulting spectra and elliptic flow on the assumed shear viscosity and initial momentum-space anisotropy.
Our work goes beyond prior viscous hydrodynamics studies by linearizing around
anisotropic background and, as a result, we are able
to better describe early-time photon production and
photon production near the transverse and longitudinal
edges of the QGP. In addition, high-momentum photon 
production is treated in a more reliable manner since
the anisotropic one-particle distribution function used
to compute the photon rates is positive definite at all
points in momentum space. Our results indicate that high-energy
photon production is quite sensitive to the
assumed level of initial momentum-space anisotropy of
the QGP. As a result, it may be possible
to experimentally constrain the early-time momentum space
anisotropy of the QGP generated in
relativistic heavy-ion collisions using the high-energy photon spectrum.

The structure of the paper is as follows. In Sec.~\ref{sec:rate}, we discuss the calculation of the photon rate from an anisotropic QGP.  In Sec.~\ref{sec:spectra}, we discuss how the rate is integrated over space-time to obtain the final differential photon yields.  In Sec.~\ref{sec:hydro}, we specify the details of the hydrodynamical evolution utilized.  In Sec.~\ref{sec:results}, we present our numerical results for different initial conditions and different values of the shear viscosity to entropy density ratio.  In Sec.~\ref{sec:conclusions}, we present a discussion of our results, conclusions, and an outlook for the future.

%%%%%%%%%%%%%%%%%%%%%%%%%%%%%%%%%%%%%%%%%%%%%%%%%%%%%%%%%%%%%%%%%%%%%%%%%%%%%%%%%%%%%%%%%%%%%%%%
\section{Photon rate in anisotropic plasma}
\label{sec:rate}
%%%%%%%%%%%%%%%%%%%%%%%%%%%%%%%%%%%%%%%%%%%%%%%%%%%%%%%%%%%%%%%%%%%%%%%%%%%%%%%%%%%%%%%%%%%%%%%%

In this section, we review the calculation of the photon production rate in an anisotropic QGP.  This method was introduced originally in Ref.~\cite{Schenke:2006yp}, but we review it here for completeness.  To proceed, one separates the contributions to the rate into those corresponding to hard-momentum and soft-momentum exchanges which we detail separately below.  In both cases, we take the quarks to be massless since, in the high-temperature and high-energy limits, the masses result in very small corrections to the relevant cross sections.  In this paper we assume that, when written in terms of LRF momentum, the one-particle distribution function is of spheroidal (Romatschke-Strickland) form~\cite{Romatschke:2003ms,Romatschke:2004jh},
\bea
f_i({\bf k}, \xi, \Lambda) = 
{f_{i}^{\rm eq}} ({\sqrt {{\bf k}^2 + \xi ({\bf k} \cdot \hat{\bf n})^2}/\Lambda)} \, ,
\label{distribution_function}
\eea
where $i = \{q, {\bar q}, g\}$, $\Lambda$ is the
transverse momentum scale, $\hat{\bf n}$ is the direction of the anisotropy, and
$\xi \ge -1$ is a parameter reflecting the strength and type of
the anisotropy.  Above $\xi$ and $\Lambda$ should be understood to be fields that depend on both space and time.  The function $f_{\rm eq}$ is an equilibrium distribution function.  In the following, we will suppress the
explicit dependence of the anisotropic distribution function on $\xi$ and $\Lambda$.
\subsubsection{Hard Contribution}
For the hard contributions, one can simply compute the Feynman diagrams corresponding to the Compton and annihilation processes.  The rate of photon production due to in-medium quark annihilation can be expressed as 
\bea
E\frac{dR_{\rm ann}}{d^3q} &=& 64\pi^3{e_q}^2\alpha_s \alpha \int_{\bf k_1} 
\frac{f_q({\bf k}_1)}{k_1} \int_{\bf k_2}\frac{f_q({\bf k}_2)}{k_2} \int_{\bf k_3} 
\frac{1+f_g({\bf k}_3)}{k_3}\nonumber\\
&& \hspace{4cm} \times \delta^4 (K_1+ K_2- Q- K_3) \Bigg [\frac{u}{t}+\frac{t}{u}\Bigg],
\label{photon_rate_ann1}
\eea
where the Mandelstam variables are defined by  $t \equiv (K_1 - Q)^2$ and 
$u \equiv (K_2 - Q)^2$, ${e_q}^2 = 2/3$,\footnote{Herein 
we include contributions from up, down, and strange quarks since these dominate the bulk of the QGP.} $\int_k \equiv \int d^3{\bf k}/(2\pi)^3$, and 
$f_{q,g}$ are the in-medium quark and gluon distribution functions.  Note that herein capital letters, e.g. $K_1$, indicate four vectors.
Henceforth, we also assume that the system is charge-conjugation symmetric such that the distribution functions 
for quarks and anti-quarks are the same, i.e. $f_q = f_{\bar q}$.

In order to regulate the IR divergence associated with
this graph, we first change variables in the
first integration to $P \equiv K_1 - Q$ and introduce an IR
cutoff $p^*$ on the integration over the exchanged
three-momentum {\bf p}~\cite{Braaten:1991dd}. 
Here we choose spherical coordinates
with the anisotropy vector $\hat{\bf n}$ defining the $z$-axis and we exploit the azimuthal symmetry around the $z$-axis to choose ${\bf q}$ to lie in the $x$-$z$ plane. Using the energy-momentum conserving delta
function and expanding out the phase-space integrals gives
\bea
E\frac{dR_{\rm ann}}{d^3q}=\frac{{e_q}^2\alpha_s \alpha}{2\pi^6}
\sum\limits_{i=1}^2 {\int_{p^*}}^{\infty} dp \, p^2 \int_{-1}^1 
d(\cos{\theta_p}) {\int_{0}}^{2\pi} d\phi_p \frac{f_q({\bf p}+{\bf q})}
{|{\bf p}+{\bf q}|} 
{\int_0}^{\infty} dk \, k \int_{-1}^1 d (\cos{\theta_k}) \nonumber\\
\times f_q({\bf k}) [1+f_g({\bf p}+ {\bf k})] \chi^{-1/2} 
\Theta(\chi)\Bigg[\frac{u}{t}\Bigg]_{\phi_{k}=\phi_i} ,
\label{photonrate_ann}
\eea
with $t = \omega^2-p^2$, $ u = (k-q)^2-({\bf k}-{\bf q})^2$, and 
$\omega = |{\bf p}+{\bf q}|-q$. The azimuthal angles $\phi_i$ are defined through
\begin{equation}
\cos({\phi_i} -{\phi_p}) = 
\frac{\omega^2-p^2+2k({\omega} - {p \cos{\theta_p} \cos{\theta_k})}}
{2pk \sin\theta_p \sin\theta_k} \, ,
\end{equation}
and $\chi \geq 0$ is given by
\begin{equation}
\chi \equiv 4 p^2k^2 \sin^2\theta_k \sin^2\theta_p
-\left[\omega^2-p^2+2k(\omega-p \cos\theta_p \cos\theta_k) \right]^2.
\end{equation}
The rate of photon production from the Compton scattering
diagrams can be obtained from
\bea
E\frac{dR_{\rm com}}{d^3q}=-128\pi^3 e_q^2\alpha_s \alpha \,
\int_{\bf k_1} \frac{f_q({\bf k}_1)}{k_1} \int_{\bf k_2}\frac{f_g({\bf k}_2)}{k_2} \int_{\bf k_3} 
\frac{1-f_q({\bf k}_3)}{k_3} \nonumber\\
\times \delta^4 (K_1+ K_2- Q- K_3) \Bigg [\frac{s}{t}+\frac{t}{s}\Bigg], 
\eea
where the Mandelstam variables 
are defined by $s \equiv (K_1 + K_2)^2$ and 
$t \equiv (K_1-Q)^2$. After changing variables
to $P \equiv K_1-Q$ and continuing as with the annihilation
process one obtains
\bea
E\frac{dR_{\rm com}}{d^3q}=-\frac{{e_q}^2\alpha_s \alpha}{2\pi^6}
\sum\limits_{i=1}^2 \int_{p^*}^{\infty} dp p^2 \int_{-1}^1 
d\cos{\theta_p} \int_{0}^{2\pi} d\phi_p \frac{f_q({\bf p}+{\bf q})}
{|{\bf p}+{\bf q}|} 
{\int_0}^{\infty} dk~ k \int_{-1}^1 d \cos{\theta_k} \nonumber\\
\times f_g({\bf k}) [1-f_q({\bf p}+ {\bf k})] \, \chi^{-1/2} \,
\Theta(\chi) \Bigg[\frac{s}{t}+\frac{t}{s}\Bigg]_{\phi_{k}=\phi_i},
\label{photon_rate_com}
\eea
with $t = {\omega^2}-p^2$ and $s= (\omega+q+k)^2 -({\bf p}+ {\bf k}+{\bf q})^2$.
The total photon production rate from hard processes is given by the sum
of Eqs.~(\ref{photonrate_ann}) and (\ref{photon_rate_com}).
\bea
E\frac{dR_{\rm hard}}{d^3q}=E\Bigg (\frac{dR_{\rm ann}}{d^3q}+\frac{dR_{\rm com}}{d^3q}\Bigg).
\label{photon_total}
\eea
We use Vegas Monte-Carlo integration \cite{Gough:2009:GSL:1538674} to evaluate the remaining five-dimensional integrals in Eqs.~(\ref{photonrate_ann}) and (\ref{photon_rate_com}). The total hard 
contribution (\ref{photon_total}) has a logarithmic infrared (IR) divergence as 
${p^*}\rightarrow 0$. This logarithmic IR divergence is cancelled by a 
corresponding ultraviolet (UV) divergence in the soft contribution which we will
describe next.

\subsubsection{Soft Contribution}

Next we calculate the contribution involving soft-momentum exchange.
We refer the reader to Ref.~\cite{Schenke:2006fz} for further details of the calculation. 
We use the Keldysh formulation, which is 
appropriate for non-equilibrium systems~\cite{Keldysh:1964ud}.
The components ($12$) and ($21$) of the polarization tensor are 
related to the emission and absorption probability of the particle
species under 
consideration, respectively~\cite{Chou:1984es, Mrowczynski:1992hq,Calzetta:1986cq}.
Due to the very low rate for photon absorbing back reactions, the rate of 
photon emission can be expressed as~\cite{Baier:1997xc} 
\begin{equation}
E\frac{dR_{\rm soft}}{d^3q}=\frac{i}{2(2\pi)^3}\left(\Pi_{12}\right)^\mu_{~\mu}\!(Q) \, , 
\label{soft_photon_rate}
\end{equation}
from the trace of the ($12$) element of the 
photon polarization tensor.
We evaluate $\left(\Pi_{12}\right)^\mu_{~\mu}$ using the HL resummed
fermion propagator derived in Ref.~\cite{Schenke:2006fz} wherein 
the authors demonstrated that the needed off-diagonal components
of the fermion self-energy can be expressed in terms of
the retarded self-energy
\begin{equation}
\Sigma (P) =\frac{C_F}{4}g^2 \int_{\bf k} \frac{f(\bf k)}{|{\bf k}|}
\frac{K \cdot \gamma}{K \cdot P} \, ,
\label{sigmap}
\end{equation}
where
\begin{equation} 
f({\bf k})\equiv 2(f_q({\bf k})+f_{\bar q}({\bf k}))+4f_g({\bf k}) \, .
\label{fk}
\end{equation}
Taking the HL limit where appropriate, one finds that 
\begin{equation}
i\left(\Pi_{12}\right)^\mu_{~\mu}\!(Q)= - e^2{e_q}^2N_c 
\frac{8f({\bf q})}{q}\int_{\bf p}Q_{\nu}{\tilde \Lambda}^{\nu} ({\bf p}) \, ,
\label{pimunu}
\end{equation}
where
\bea
{\tilde\Lambda}^{\nu} ({\bf p})= [ {\Lambda^{\nu\alpha}}_\alpha (P)
- {\Lambda_\alpha}^{\nu\alpha}(P)+ {\Lambda_\alpha}^{\alpha\nu}(P)]_{p_0=p(\hat {\bf p} \cdot \hat {\bf q})} \, .
\label{soft5}
\eea
The tensor $\Lambda$ is defined through
\begin{equation}
\Lambda_{\alpha \beta\gamma}(P)=\frac{P_{\alpha}-\Sigma_{\alpha}(P)}
{(P-\Sigma(P))^2} \, {\rm Im}[\Sigma_{\beta}(P)] \, \frac{P_{\gamma}-\Sigma_\gamma^*(P)}
{(P-{\Sigma^*(P)})^2} \, ,
\label{soft6}
\end{equation}
where a star indicates complex conjugation. To compute the soft photon 
rate~(\ref{soft_photon_rate}), we numerically evaluate Eqs.~(\ref{sigmap}) 
and~(\ref{pimunu}) with an UV cutoff $p^*$ placed on the length 
of the three-momentum in Eq.~(\ref{pimunu}).

%%%%%%%%%%%%%%%%%%%%%%%%%%%%%%%%%%%%%%%%%%%%%%%%%%%%%%%%%%%%%%%%%%%%%%%%%%%%%%%%%%%%%
\begin{figure*}[t]
\centerline{\includegraphics[width=0.7\linewidth]{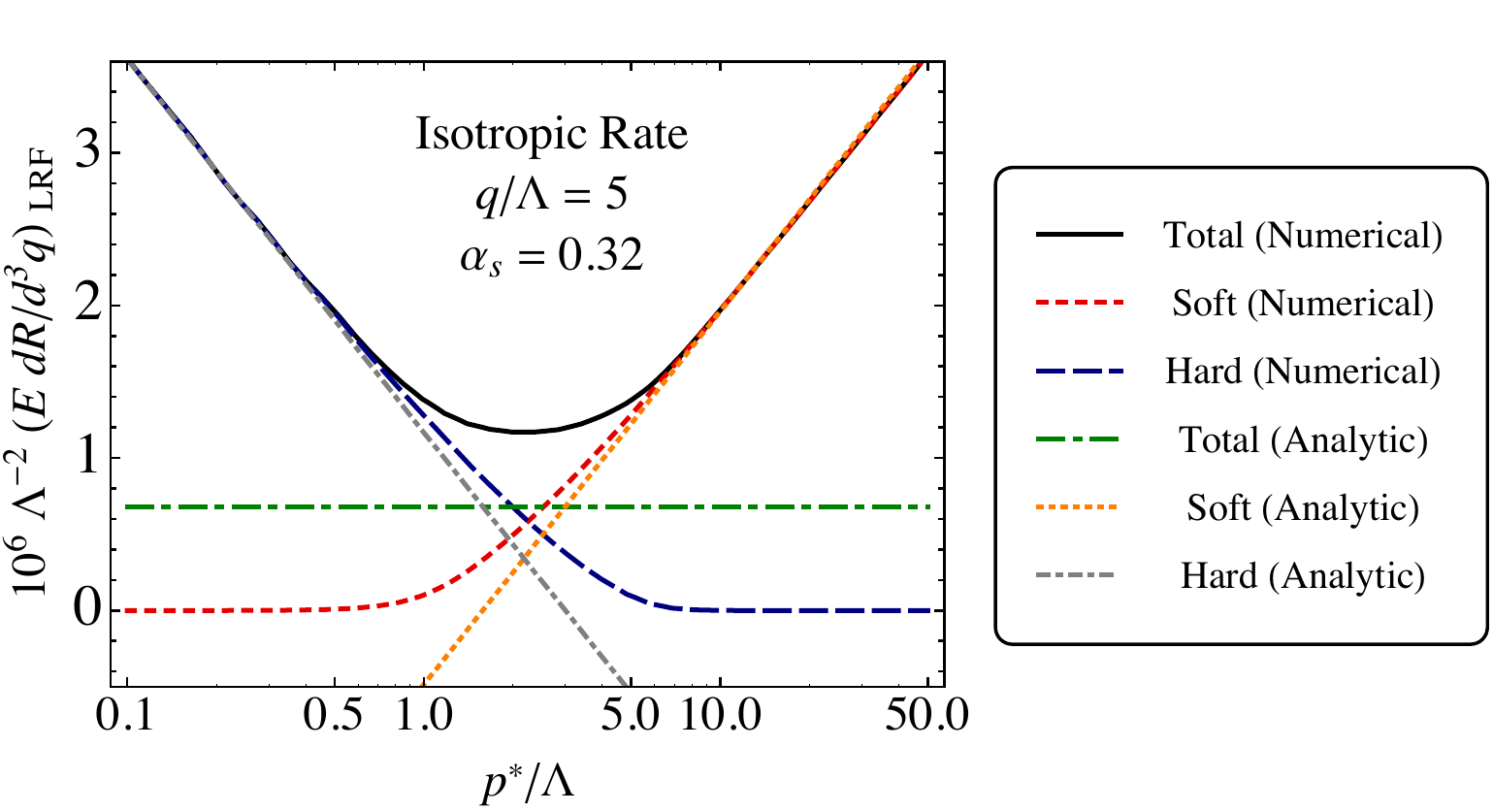}}
\caption{Dependence of the isotropic photon rate on the separation scale $p^*$.  The various curves show the total rate determined numerically, the soft and hard rates determined numerically, the LO rate analytic result of Braaten and Yuan \cite{Braaten:1991dd}, and the separate Braaten and Yuan results for the soft and hard rates.}
\label{fig:pstarFig}
\end{figure*}
%%%%%%%%%%%%%%%%%%%%%%%%%%%%%%%%%%%%%%%%%%%%%%%%%%%%%%%%%%%%%%%%%%%%%%%%%%%%%%%%%%%%%%%%%%%%%%%

In Fig.~\ref{fig:pstarFig} we show the soft and hard contributions to the isotropic ($\xi=0$) photon rate and the total rate obtained by summing these contributions.  We compare our numerical results with the analytic estimates of Braaten and Yuan \cite{Braaten:1991dd} for both the individual contributions and the total rate.  For this figure, we consider the case of a realistic coupling with $\alpha_s=0.32$ and a photon momentum which is five times the corresponding transverse momentum scale $\Lambda$.\footnote{Since $\xi=0$, $\Lambda$ can be identified with the temperature $T$ in this figure, but we have kept the labels general for comparison with other figures.}  As can be seen from Fig.~\ref{fig:pstarFig}, there is a logarithmic IR divergence at small $p^*/\Lambda$ and logarithmic UV divergence at large $p^*/\Lambda$.   However, even for large $\alpha_s$, there is a window over which the result does not depend strongly on the choice of $p^*$.  In practice, we use the \textit{principle of minimal sensitivity} (PMS) to set $p^*$ by requiring the derivative of the rate with respect to $p^*$ to vanish (minimum of the black solid curve in Fig.~\ref{fig:pstarFig}).  The other things we see from this figure are that the analytic calculations do well in capturing the asymptotic regimes of each contribution, but that the total result obtained from analytic method is lower by approximately a factor of 2 due to the fact that the analytic approximations made to obtain the Braaten-Yuan rate fail to accurately reproduce the numerically integrated rate.\footnote{If one takes very small $\alpha_s$, e.g. $\alpha_s = 0.01$, then the numerical and analytic results agree nearly perfectly for large $q/\Lambda$, giving us confidence in our numerical methods.}

\begin{figure*}[t]
\centerline{\includegraphics[width=0.97\linewidth]{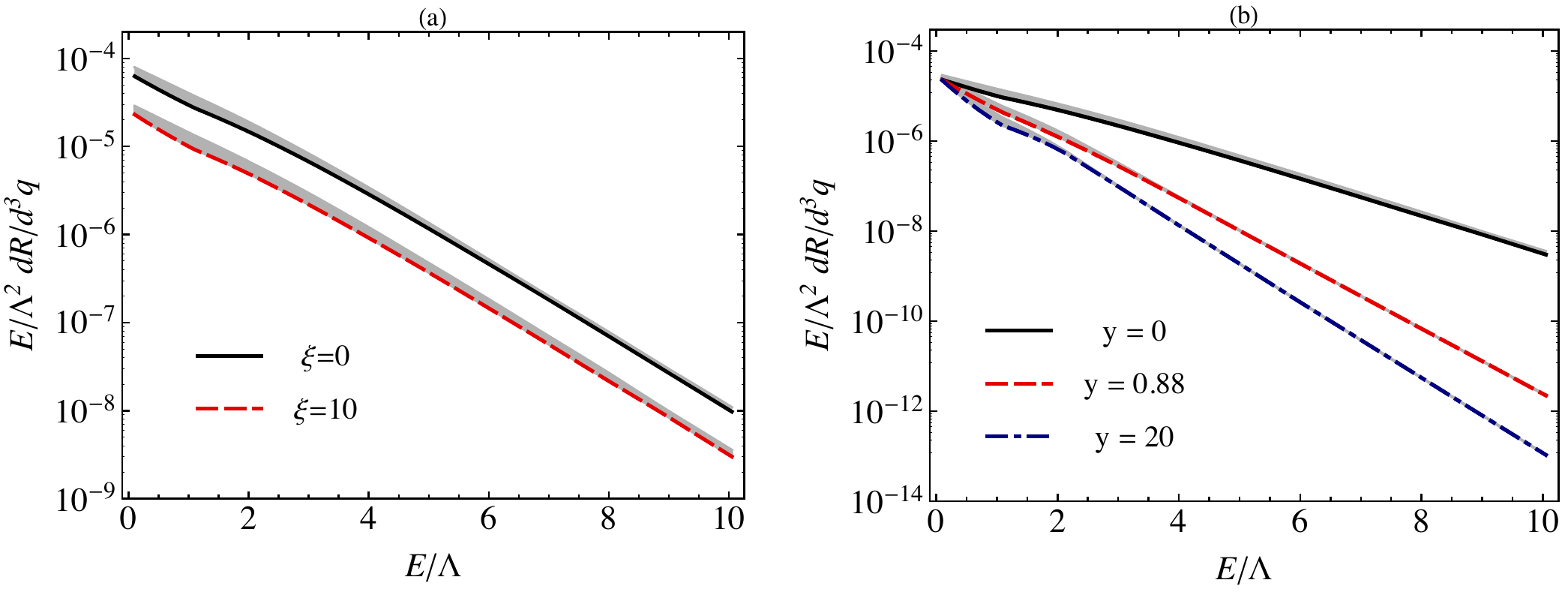}}
\caption{Rate as a function of the scaled photon energy for (a) $\xi \in \{0,10\}$ for $y=0$ and (b) $\xi = 10$ for $y \in \{0,0.88,20\}$.  The shaded bands show the change in the result obtained by varying the separation scale $p^* \rightarrow 2p^*$.}
\label{fig:rateFig}
\end{figure*}

In Fig.~\ref{fig:rateFig} we plot the photon rate as a function of the scaled photon energy $E/\Lambda$ for (a) $\xi \in \{0,10\}$ for $y=0$ and (b) $\xi=10$ for $y \in \{0,0.88,20\}$.  In both panels of Fig.~\ref{fig:rateFig}, the lines are the result of evaluating the rate at the PMS value of $p^*$ and the shaded regions indicate the variation of the result obtained when varying $p^* \rightarrow 2 p^*$.  In the high-energy and weak-coupling limit, the dependence on $p^*$ formally vanishes, however, for realistic couplings, direct numerical evaluation of the integrals defining the rate allows us to gauge this uncertainty.  From the results we find that there is a $\lesssim$ 30\% variation of the photon rate at $E/\Lambda=1$ and a $\lesssim$ 20\% variation at $E/\Lambda=10$.  From  Fig.~\ref{fig:rateFig}(b) we conclude that there is a significant rapidity dependence of the photon rate when the system is momentum-space anisotropic, with production peaked at mid-rapidities.  This is simply due to the dominance of forward scattering coupled to production from a momentum-space anisotropic source with $\langle p_T^2 \rangle > \langle p_L^2 \rangle$.  Note, however, that, if the system were exactly boost-invariant then, when integrated over all spatial rapidity, the final photon spectrum would not depend on the photon rapidity.  If, on the other hand, the system is not boost-invariant, then we expect to see a suppression of photon production at forward/backward rapidities if the LRF distribution is oblate.

%%%%%%%%%%%%%%%%%%%%%%%%%%%%%%%%%%%%%%%%%%%%%%%%%%%%%%%%%%%%%%%%%%%%%%%%%%%%%%%%%%%%%
%%%%%%%%%%%%%%%%%%%%%%%%%%%%%%%%%%%%%%%%%%%%%%%%%%%%%%%%%%%%%%%%%%%%%%%%%%%%%%%%%%%%%%%%%%%%%%%

%%%%%%%%%%%%%%%%%%%%%%%%%%%%%%%%%%%%%%%%%%%%%%%%%%%%%%%%%%%%%%%%%%%%%%%%%%%%%%%%%%%%%%%%%%%%%%%%
\section {Photon spectrum}
\label{sec:spectra}
%%%%%%%%%%%%%%%%%%%%%%%%%%%%%%%%%%%%%%%%%%%%%%%%%%%%%%%%%%%%%%%%%%%%%%%%%%%%%%%%%%%%%%%%%%%%%%%%

As mentioned previously, in this paper we want to study the impact of space-time dependent anisotropies on the photon differential spectrum at high transverse momentum. In this way we can probe the early stages of the QGP, when the anisotropies are expected to be the largest.  In order to accomplish this, we need to convolute Eqs.~(\ref{photon_total}) and~(\ref{soft_photon_rate}) with the space-time dependence of $\Lambda$ and $\xi$ to integrate the rate over the QGP space-time volume and obtain the final spectrum of photons emitted from the deconfined QGP over its lifetime.  We now discuss how this is implemented.

The photon four-momentum is parametrized in the standard manner 
\begin{equation}
q^{\mu}=q_{\perp} (\cosh y, \cos \phi_q, \sin \phi_q, \sinh y) \, ,
\label{mompar}
\end{equation}
where $y \equiv \ln[(E+q_\parallel)/(E-q_\parallel)]/2$ is the momentum-space rapidity. Above $q_{\perp}$, $q_{\parallel}$, and $\phi_q$ are the transverse momentum, longitudinal momentum, and momentum azimuthal angle, respectively.  For space-time, we use the usual Milne hyperbolic parametrization for describing  heavy-ion collisions within the relativistic hydrodynamics framework
\begin{equation}
x^{\mu}=(\tau \cosh \varsigma, {\bf x}_\perp, \tau \sinh \varsigma) \, . 
\label{spacepar}
\end{equation}
In Eq.~(\ref{spacepar}), $\tau\equiv\sqrt{t^2 - z^2}$ is the longitudinal proper time, ${\bf x}_\perp$ is a two-vector containing the transverse coordinates, and $\varsigma \equiv \tanh^{-1} (z/t)$ is the space-time rapidity. With these parametrizations, the differential measures for four-momentum  and space-time are $d^4 q= E \,dy \,q_{\perp} dq_\perp \,d\phi_q$ and $d^4 x=\tau \, d\tau \,d\varsigma \, d^2 x_\perp$, respectively. This allows us to calculate the differential spectrum using
\begin{equation}
\frac{dN}{q_\perp dq_\perp dy}
=\int_{0}^{2\pi} \!\! d\phi_q \int \! d^4x \left( \! E\frac{dR}{d^3q} \right)_{\rm LRF} . 
\label{pTspectrum}
\end{equation}
The integration over the space-time volume is performed solely in the 
deconfined QGP stage. We only include contributions from regions that have an effective temperature higher than a critical temperature, 
i.e. $T_{\rm eff} \equiv {\cal R}^{1/4}(\xi)\Lambda > T_c$ with ${\cal R}(\xi)$ defined in Eq.~(\ref{R}).  In all results 
shown herein, we assume $T_c = 175$ MeV.  We will gauge the sensitivity of our results to this assumption in Sec.~\ref{sec:results}.  We will assume that when the system reaches $T_c$, all QGP medium emission stops. 
We do not consider the emission from the mixed/hadronic phase herein. 

Finally, we mention that we evaluate the left-hand-side of Eq.~(\ref{pTspectrum}) in the center of mass of the colliding nuclei (LAB frame), while the photon rate is calculated in the LRF of the emitting region.  Therefore, before evaluating Eq.~(\ref{pTspectrum}), we have to boost the LAB frame momentum $q^{\mu}$ to the LRF of the fluid cell using a Lorentz boost $q^\mu_{\rm LRF} = \Lambda^{\mu\,\,}_{\,\,\nu} q^\nu$,  where the Lorentz boost tensor
\begin{equation}
\Lambda^{\mu\,\,}_{\,\,\nu}(u^{\lambda}) \equiv \left(
\begin{array}{rrrr}
\gamma &  -\gamma v_x  &  -\gamma v_y  &  -\gamma v_z \\
-\gamma v_x & 1 + (\gamma - 1) \frac{v_x^2}{v^2}    &     (\gamma - 1) \frac{v_x v_y}{v^2}  &     (\gamma - 1) \frac{v_x v_z}{v^2} \\
-\gamma v_y &     (\gamma - 1) \frac{v_x v_y}{v^2}  & 1 + (\gamma - 1) \frac{v_y^2}{v^2}    &     (\gamma - 1) \frac{v_y v_z}{v^2} \\
-\gamma v_z &     (\gamma - 1) \frac{v_x v_z}{v^2}  &     (\gamma - 1) \frac{v_y v_z}{v^2}  & 1 + (\gamma - 1) \frac{v_z^2}{v^2}
\end{array} \right),
\label{boost}
\end{equation}
depends on the four-velocity of the fluid element $u^{\mu}(x^{\lambda}) \equiv \gamma (1, v_x, v_y, v_z)$, with $\gamma \equiv 1 / \sqrt{1 - v^2}$ and $v \equiv \sqrt{v_z^2 + v_y^2 +v_z^2}$.  Making use of Eqs.~(\ref{photon_total}) and (\ref{soft_photon_rate}) in Eq.~(\ref{pTspectrum}), we obtain the photon spectrum including the effect of a space-time-dependent momentum anisotropy and taking into account the effect of the dynamically-generated collective flow of the QGP.

%%%%%%%%%%%%%%%%%%%%%%%%%%%%%%%%%%%%%%%%%%%%%%%%%%%%%%%%%%%%%%%%%%%%%%%%%%%%%%%%%%%%%%%%%%%%%%%%
\section{(3+1)D anisotropic hydrodynamics}
\label{sec:hydro}
%%%%%%%%%%%%%%%%%%%%%%%%%%%%%%%%%%%%%%%%%%%%%%%%%%%%%%%%%%%%%%%%%%%%%%%%%%%%%%%%%%%%%%%%%%%%%%%%

As mentioned above, in order to obtain the predictions for the differential photon spectrum expected to be produced from the QGP phase, we must integrate over the full space-time history of the QGP. For this purpose we use a (3+1)D leading-order spheroidal anisotropic hydrodynamics (aHydro) code~\cite{Florkowski:2010cf,Martinez:2010sc,Ryblewski:2010bs,Martinez:2010sd,Ryblewski:2011aq,Martinez:2012tu,Ryblewski:2012rr}.  The aHydro framework reduces to second-order viscous hydrodynamics in the limit of small anisotropy \cite{Tinti:2014yya}, but reproduces the dynamics of the QGP more reliably when there are large momentum-space anisotropies.

We assume that the QGP created during the collision of the heavy ions evolves through a non-equilibrium state and that the quark and anti-quark one-particle distribution functions are well-approximated by a time-evolving distribution of the form specified in Eq.~(\ref{distribution_function}) both at early times and late times. At the same time, we also assume that, although the system is highly anisotropic, it may still be, to good approximation, described using hydrodynamic-like degrees of freedom, such as energy density and pressures.\footnote{This assumption has been tested elsewhere by comparing the predictions of anisotropic hydrodynamics to exact solutions of the Boltzmann equation in a variety of special cases \cite{Florkowski:2013lza,Florkowski:2013lya,Bazow:2013ifa,Florkowski:2014sfa,Florkowski:2014sda,Denicol:2014xca,Denicol:2014tha,Nopoush:2014qba}.}   As a result, the detailed microscopic description of the system can be replaced by an effective description written in terms of simple physical laws, such as conservation of energy and momentum.

The equations of motion for the anisotropic system are obtained by starting from kinetic theory, assuming that the form of the distribution function of the system is known at the leading order and given by the form (\ref{distribution_function}). This can be performed by taking moments of the Boltzmann kinetic equation.  As usual, the collisional kernel must be specified in the Boltzmann kinetic equation.  Here we will assume that the collisional kernel can be treated in the relaxation-time approximation (RTA) such that
\begin{equation}
p^\mu \partial_\mu f = -\frac{p^\mu u_\mu}{\tau_{\rm eq}}(f-f_{\rm eq})\,,
\label{eq:boltzmanneq}
\end{equation}
where $\tau_{\rm eq}$ is the microscopic relaxation time which can depend on position and time.
Taking the first moment of the Boltzmann equation results in the energy-momentum conservation equation
\begin{equation}
\partial_\mu T^{\mu \nu} = 0 \, .
\label{enmomcon}
\end{equation}
Taking the zeroth moment of the Boltzmann equation results in the particle production equation
\begin{equation}
\partial_\mu j^{\mu} = -u_\mu \frac{j^{\mu}-j^{\mu}_{\rm eq}}{\tau_{\rm eq}} \, .
\label{partprod}
\end{equation}
At leading order, the aHydro energy-momentum tensor has the form typical for a spheroidally anisotropic system
\begin{equation}
T^{\mu \nu} = \left( \varepsilon  + P_{\perp}\right) u^{\mu}u^{\nu} - P_{\perp} \, g^{\mu\nu} - (P_{\perp} - P_{\parallel}) z^{\mu}z^{\nu} \, ,
\label{Taniso}
\end{equation}
and the particle flux is defined in the standard manner\footnote{We assume vanishing chemical potential gradients.}
\begin{equation}
j^{\mu} =  n \, u^\mu \, .
\label{Naniso}
\end{equation}
In Eqs.~(\ref{Taniso}) and (\ref{Naniso}) $\varepsilon$, $n$, $P_{\parallel}$, and $P_{\perp}$ stand for energy density, particle density, longitudinal pressure, and transverse pressure, respectively. The four-vector $z^\mu$ is orthogonal to $u^{\mu}$ and in the LRF points in the longitudinal direction (identified with the direction of the dynamically-evolving anisotropy in the system, $\bf \hat n$)~\cite{Martinez:2012tu}.

Equations (\ref{enmomcon}) and (\ref{partprod}) provide a set of five independent partial differential equations
\begin{eqnarray}
D_u \varepsilon &=& - \left( \varepsilon+P_\perp \right) \theta_u 
+ \left( P_\perp-P_\parallel \right) u_\nu  D_z z^\nu \, , \hspace{6mm}
\label{enmomconU} \\
D_z P_\parallel &=&  \left( P_\perp-P_\parallel \right) \theta_z  + \left( \varepsilon+ P_\perp \right) z_\nu  D_u u^\nu \, , 
\label{enmomconeqs}\\
D_u u_{\perp} &=& - \frac{u_{\perp}}{\varepsilon + P_{\perp}} \Bigg[ \frac{ {\bf u}_\perp \cdot {\boldsymbol \nabla}_\perp P_{\perp}}{u_\perp^2} 
% \nonumber \\ & & \hspace{7mm} 
+  D_u P_{\perp} +  (P_{\perp} - P_{\parallel}) u_\nu D_z z^\nu \frac{}{} \Bigg] ,
\label{HydEqEuler1}\\
D_u \left( \frac{u_x}{u_y} \right) &=& \frac{1}{u_y^2 (\varepsilon + P_{\perp})} \left( u_x \partial_y - u_y \partial_x \right)P_{\perp} \, ,
\label{HydEqEuler2}
\end{eqnarray}
and 
\begin{equation}
\frac{D_u \xi}{2 (1+\xi)} - \frac{3 D_u \Lambda}{\Lambda} = \theta_u + \frac{1}{\tau_{\rm eq}} \left[1 - {\cal R}^{3/4}(\xi)\sqrt{1+\xi} \right] ,
\label{partprodeqs}
\end{equation}
respectively, for five parameters:  the three independent components of the four-velocity $u^\mu$, the transverse momentum scale $\Lambda$, and the anisotropy parameter $\xi$. In the above equations, we use boldfaced letters with a $\perp$ subscript to indicate two-dimensional vectors in the transverse plane, e.g. ${\bf u_\perp} \equiv (u_x,u_y)$ and ${\boldsymbol\nabla}_\perp \equiv (\partial_x,\partial_y)$.   We have also introduced a compact notation for the convective derivative $D_u \equiv u^\mu \partial_\mu$, the longitudinal derivative $D_z \equiv z^\mu \partial_\mu$, and the expansion scalars $\theta_u \equiv \partial_\mu u^\mu$ and $\theta_z \equiv \partial_\mu z^\mu$.

We use the following parametrizations of the LAB frame four-velocity of the fluid $u^\mu$ 
and the space-like four-vector $z^\mu$
\begin{eqnarray}
u^\mu &=& (u_0 \cosh \vartheta, {\bf u_\perp}, u_0 \sinh \vartheta) \,  , \label{U3+1} \\
z^\mu &=& (	 \sinh \vartheta, {\bf 0},  \cosh \vartheta) \, , \label{V3+1}
\end{eqnarray}
where we introduced the longitudinal rapidity of the fluid cell $\vartheta$. Using the four-velocity normalization 
condition, $u^\mu u_\mu =1$, one has
\begin{eqnarray}
u_0 &=& \sqrt{1+u_\perp^2} \, , \nonumber \\
u_\perp &\equiv& \sqrt{u_x^2 + u_y^2} \, .
\label{u0}
\end{eqnarray}
With the parametrizations (\ref{U3+1}) and (\ref{V3+1}), one may calculate the following quantities appearing in 
Eqs.~(\ref{enmomconU})-(\ref{partprodeqs}),
\begin{eqnarray}
D_u &=& {\bf u}_\perp \cdot {\bf \nabla}_\perp + u_0 \hat{L}_1 \, , \\
\theta_u &=& {\bf \nabla}_\perp \cdot {\bf u}_\perp + \hat{L}_1 u_0 + u_0 \hat{L}_2 \vartheta \, , \\
D_z &=&  \hat{L}_2 \, , \\
\theta_z &=&  \hat{L}_1 \vartheta \, , \\
u_\nu D_z z^\nu &=& u_0 \hat{L}_2 \vartheta \, , \\
z_\nu D_u u^\nu &=& - u_0 \left( {\bf u}_\perp \cdot 
{\bf \nabla}_\perp + u_0 \hat{L}_1  \right) \vartheta \, ,
\label{op1}
\end{eqnarray}
where the two linear differential operators, $\hat{L}_1$ and $\hat{L}_2$, are given by
\begin{eqnarray}
\hat{L}_1  &=& \cosh (\varsigma - \vartheta) \partial_\tau - \sinh (\varsigma - \vartheta) \frac{\partial_\varsigma}{\tau} \, , \label{op21}\\
-\hat{L}_2  &=& \sinh (\varsigma - \vartheta) \partial_\tau - \cosh (\varsigma - \vartheta) \frac{\partial_\varsigma}{\tau} \, . \label{op22}
\end{eqnarray}
We also use the relation between the relaxation time $\tau_{\rm eq}$ and the shear viscosity to entropy 
density ratio $\bar{\eta} \equiv \eta /s$ \cite{Martinez:2010sc},
\begin{eqnarray}
\tau_{\rm eq} = \frac{5 \bar{\eta}}{2 T} \, .
\label{reltime}
\end{eqnarray}

\subsection{Anisotropic equation of state}
\label{ssec:eos}
%%%%%%%%%%%%%%%%%%%%%%%%%%%%%%%%%%%%%%%%%%%%%%%%%%%%%%%%%%%%%%%%%%%%%%%%%%%%%%%%%%%%%%%%%%%%%%%%

In this paper we consider a system that consists of massless particles described by the anisotropic distribution 
function (\ref{distribution_function}). Using standard kinetic theory definitions
\begin{eqnarray}
N^\mu &\equiv& \int dK \,k^\mu f  \, , \\
T^{\mu\nu} &\equiv& \int dK \,k^\mu k^\nu f  \, ,
\end{eqnarray}
where $dK \equiv d^3{\bf k} / \left[ (2 \pi)^3 k^0 \right]$, and the tensor decompositions specified in 
Eqs.~(\ref{Taniso}) and (\ref{Naniso}), one can calculate the thermodynamic properties of the system \cite{Martinez:2009ry}
\begin{eqnarray}
\label{densaniso}
n(\Lambda, \xi) &=& \frac{n_{\rm iso}(\Lambda)}{\sqrt{1+\xi}}\, ,\\
\label{energyaniso}
{\cal \varepsilon}(\Lambda, \xi) &=& {\cal R}(\xi)\,\varepsilon_{\rm iso}(\Lambda)\, ,\\
\label{transpressaniso}
P_\perp(\Lambda, \xi) &=& {\cal R}_\perp(\xi)\,P_{\rm iso}(\Lambda)\, ,\\
\label{longpressaniso}
P_\parallel(\Lambda, \xi) &=& {\cal R}_\parallel(\xi)\,P_{\rm iso}(\Lambda)\, ,
\end{eqnarray}
where $n_{\rm iso}$, $\varepsilon_{\rm iso}$, and $P_{\rm iso}$ are the isotropic particle density, 
energy density, and pressure, respectively, and
\begin{eqnarray}
\label{R}
{\cal R}(\xi) &\equiv& \frac{1}{2}\left[\frac{1}{1+\xi}
+\frac{\tan^{-1}\sqrt{\xi}}{\sqrt{\xi}} \right] ,  \\ 
\label{RT}
{\cal R}_\perp(\xi) &\equiv& \frac{3}{2 \xi} 
\left[ \frac{1+(\xi^2-1){\cal R}(\xi)}{\xi + 1}\right]
 \, , 
\\ 
\label{RL}
{\cal R}_\parallel(\xi) &\equiv&  \frac{3}{\xi} 
\left[ \frac{(\xi+1){\cal R}(\xi)-1}{\xi+1}\right] .
\end{eqnarray}
Herein, we assume the simple case of a conformal ideal fluid, i.e. 
$\varepsilon_{\rm iso} = 3 P_{\rm iso}$. As a result, 
Eqs.~({\ref{densaniso}})--({\ref{longpressaniso}}) describe the 
equation of state of an anisotropic system of classical massless particles 
with vanishing chemical potential.
%

%%%%%%%%%%%%%%%%%%%%%%%%%%%%%%%%%%%%%%%%%%%%%%%%%%%%%%%%%%%%%%%%%%%%%%%%%%%%%%%%%%%%%%%%%%%%%%%%
\subsection{Initial conditions}
\label{ssec:ini}
%%%%%%%%%%%%%%%%%%%%%%%%%%%%%%%%%%%%%%%%%%%%%%%%%%%%%%%%%%%%%%%%%%%%%%%%%%%%%%%%%%%%%%%%%%%%%%%%

In order to solve the set of partial differential equations (\ref{enmomconU})--(\ref{partprodeqs}), one has to specify the initial conditions at the initial longitudinal proper-time for the hydrodynamic evolution, ${\tau = \tau_0}$, i.e. one has to define five three-dimensional profiles: $\Lambda (\tau_0, {\bf x_\perp}, \varsigma)$, $\xi (\tau_0, {\bf x_\perp}, \varsigma)$, $u_x (\tau_0, {\bf x_\perp}, \varsigma)$, $u_y (\tau_0, {\bf x_\perp}, \varsigma)$, and  $\vartheta (\tau_0, {\bf x_\perp}, \varsigma)$.  During the initial moments of a heavy-ion collision, due to inelastic interactions, the participating nucleons deposit energy into the space-time volume of the fireball. In this work, we assume that the distribution of deposited energy is well described by the optical Glauber model.  As a result, the transverse momentum scale is given by 
\begin{equation}
\Lambda (\tau_0, {\bf x_\perp}, \varsigma) = 
\varepsilon^{-1}_{\rm iso}\!\left( \varepsilon_0 \frac{\rho(b, {\bf x_\perp}, \varsigma)}{\rho(0, {\bf 0}, 0)} \right) ,
\label{inilambda}
\end{equation}
where the proportionality constant $\varepsilon_0$ is chosen in such a way as to reproduce the total number of charged particles measured in the experiment, and $\varepsilon^{-1}_{\rm iso}$ denotes the inverse function of $\varepsilon_{\rm iso}(\Lambda)$.  For a central collision, in what follows, we identify $\Lambda_0$ as the initial transverse momentum scale at the center of the simulated region.

The density of sources is constructed using a standard mixed model
\begin{equation}
\rho(b, {\bf x_\perp}, \varsigma) \equiv  \left[\!\frac{}{} (1 - \kappa) (\rho_{\rm WN}^+(b, {\bf x_\perp}) +\rho_{\rm WN}^-(b, {\bf x_\perp})) 
+ \, 2 \,\kappa\, \rho_{\rm BC}(b, {\bf x_\perp}) \frac{}{}\!\right] h(\varsigma - \varsigma_S({b, \bf x_\perp})) \, , \;\;
\label{sources}
\end{equation}
where $\rho^{\pm}_{\rm WN}$ is the density of wounded nucleons from the left/right-moving nuclei and $\rho_{\rm BC}$ is the density of binary collisions, both of which are obtained using the optical limit of the Glauber model
\begin{eqnarray}
\rho_{\rm WN}^\pm(b, {\bf x_\perp})  &\equiv& T\left( {\bf x_\perp}\!\mp\!\frac{{\bf b_\perp}}{2}\right)\!\left[ 1\! -\! e^{- \sigma_{NN} T\left( {\bf x_\perp} \pm  \frac{{\bf b_\perp}}{2}\right)} \right], \hspace{5mm} \label{sourcesGlauber1}\\ 
\rho_{\rm BC}(b, {\bf x_\perp})  &\equiv& \sigma_{NN}T\left( {\bf x_\perp}\!+\! \frac{{\bf b_\perp}}{2}\right) T\left( {\bf x_\perp}\!-\!\frac{{\bf b_\perp}}{2}\right) .
\label{sourcesGlauber2}
\end{eqnarray}
The longitudinal profile is taken to be~\cite{Bozek:2012qs}
\begin{eqnarray}
h(\varsigma) \equiv \exp \left[ - \frac{(\varsigma - \Delta \varsigma)^2}{2 \sigma_\varsigma^2} \Theta (|\varsigma| - \Delta \varsigma) \right] .
\label{longprof}
\end{eqnarray}

For the LHC case studied here, we use $\kappa = 0.145$ for the mixing factor and an inelastic cross-section $\sigma_{\rm NN} = 62$ mb. The parameters of the longitudinal profile (\ref{longprof}) were fitted to reproduce the pseudorapidity distribution of charged particles with the results being $\Delta\varsigma = 2.5$ and $\sigma_{\varsigma} = 1.4$. The shift in rapidity is calculated according to the formula \cite{Bozek:2009ty}
\begin{eqnarray}
\varsigma_S \equiv \frac{1}{2}\ln \frac{\rho_{\rm WN}^++ \rho_{\rm WN}^- 
+ v_P (\rho_{\rm WN}^+- \rho_{\rm WN}^-)}{\rho_{\rm WN}^++ \rho_{\rm WN}^- - v_P (\rho_{\rm WN}^+- \rho_{\rm WN}^-)} \, ,
\label{shift}
\end{eqnarray}
where all functions are understood to be evaluated at a particular value of $b$ and ${\bf x_\perp}$. The participant velocity is defined as $v_P \equiv \sqrt{(\sqrt{s}/2)^2-(m_N/2)^2}/(\sqrt{s}/2)$ and $m_N$ is the nucleon mass.  In Eqs.~(\ref{sourcesGlauber1})--(\ref{sourcesGlauber2}) we have made use of the thickness function
\begin{eqnarray}
T({\bf x_\perp}) \equiv \int dz \,\rho_{\rm WS}({\bf x_\perp},z) \, ,
\label{thickness}
\end{eqnarray}
where the nuclear density is given by the Woods-Saxon profile
\begin{equation}
\rho_{\rm WS}({\bf x_\perp},z) \equiv \rho_0 \left[ 1 + 
\exp\left(\frac{\sqrt{{\bf x_\perp}^2 + z^2} - R}{a}\right)\right]^{-1}.
\label{WS}
\end{equation}
For Pb-Pb collisions, we use $\rho_0 = 0.17\, {\rm fm}^{-3}$ for the nuclear saturation density, $R = 6.48$ fm for the nuclear radius, and $a = 0.535$ fm for the surface diffuseness of the nucleus. 

%%%%%%%%%%%%%%%%%%%%%%%%%%%%%%%%%%%%%%%%%%%%%%%%%%%%%%%%%%%%%%%%%%%%%%%%%%%%%%%%%%%%%%%%%%%%%%%%
\section{Results}
\label{sec:results}
%%%%%%%%%%%%%%%%%%%%%%%%%%%%%%%%%%%%%%%%%%%%%%%%%%%%%%%%%%%%%%%%%%%%%%%%%%%%%%%%%%%%%%%%%%%%%%%%

We now present our results for the photon spectrum and elliptic flow of QGP-generated photons.  We focus here on Pb-Pb collisions with nucleon-nucleon center of mass energies of $\sqrt{s_{\rm NN}} =$ 2.76 TeV.  Since the hard and soft contribution of differential photon rate described by Eqs.~(\ref{photon_total}) and (\ref{soft_photon_rate}) are independent of the model assumed for the space-time evolution of the system, we first numerically compute the dimensionless differential photon rate $E/\Lambda^2 dR/d^3q$ on a uniformly-spaced three-dimensional grid in ${0.1 \leq q_\perp/\Lambda \leq 30}$, ${0 \leq |y| \leq 10}$, and ${-1 \leq {\rm log}_{10}(\xi+1) \leq 2.5}$. The grid spacings used for these three variables were chosen in such a way that the full three-dimensional function is well approximated at continuous values using a spline-based interpolating function.  For the final integration, we use the Vegas Monte-Carlo method to numerically integrate Eq.~(\ref{pTspectrum}) over space-time and transverse momentum angle.\footnote{During integration we set the rate to zero outside of the interpolated region.  The excluded regions give a negligible contribution to the integrated rate.}

In all plots we assume a min-bias collision with $b=9.5$ fm and we begin the aHydro evolution at $\tau_0 = 0.3$ fm/c.  At $\tau_0$, we assume that the produced matter has no transverse flow, i.e. $u_x (\tau_0, {\bf x_\perp}, \varsigma) = u_x (\tau_0, {\bf x_\perp}, \varsigma) = 0$, while the initial longitudinal flow is of Bjorken form $\vartheta (\tau_0, {\bf x_\perp}, \varsigma) = \varsigma$.  We also assume that the initial anisotropy field is homogeneous, $\xi (\tau_0, {\bf x_\perp}, \varsigma) = \xi_0$.  Finally, the initial central transverse momentum scale $\Lambda_0$ used for all results is specified in Table \ref{table:edensities}.  These values were tuned by requiring the final particle multiplicity to be a constant as $\eta/s$ and $\xi_0$ were varied.  In all plots included herein we show results for central rapidity ($y=0$) and have used a fixed $\alpha_s = 0.32$. 

%%%%%%%%%%%%%%%%%%%%%%%%%%%%%%%%%%%%%%%%%%%%%%%%%%%%%%%%%%%%%%%%%%%%%%%%%%%%%%%%%%%%%
\begin{table}[t]
\begin{center}
\begin{tabular}{|c|c|c|c|}
\hline
\diaghead{\theadfont aaaaaaaaaaaaaaaa}%
{\hspace{3mm}$4 \pi \eta/s$}{$\xi_0$\hspace{3mm}} & ~0~ & ~10~ & ~100~ \\   
\hline \hline 1 & ~0.552 & ~0.765~ & 1.009~ \\ 
\hline 2 & ~0.546~ & ~0.752~ & ~0.992~ \\
\hline 3 & ~0.544~ & ~0.748~ & ~0.990~ \\
\hline 
\end{tabular}
\end{center}
\caption{\small Values of the initial transverse momentum scale $\Lambda_{\rm 0}$ in GeV used in all figures in the results section.}
\label{table:edensities}
\end{table}
%%%%%%%%%%%%%%%%%%%%%%%%%%%%%%%%%%%%%%%%%%%%%%%%%%%%%%%%%%%%%%%%%%%%%%%%%%%%%%%%%%%%%

%%%%%%%%%%%%%%%%%%%%%%%%%%%%%%%%%%%%%%%%%%%%%%%%%%%%%%%%%%%%%%%%%%%%%%%%%%%%%%%%%%%%%
\begin{figure*}[t]
\hspace{2mm}
\includegraphics[width=0.5\linewidth]{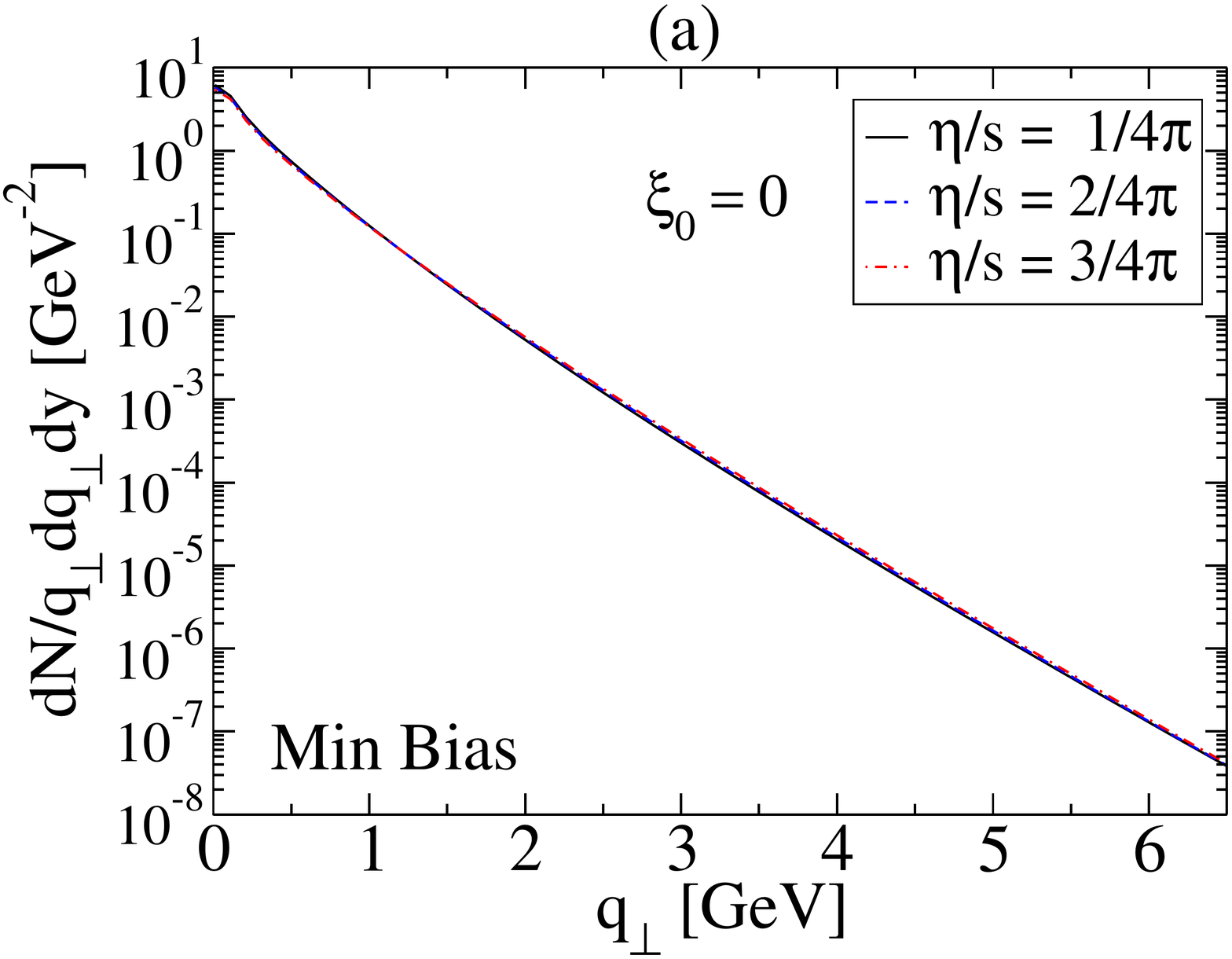}\hspace{-5mm}
\includegraphics[width=0.5\linewidth]{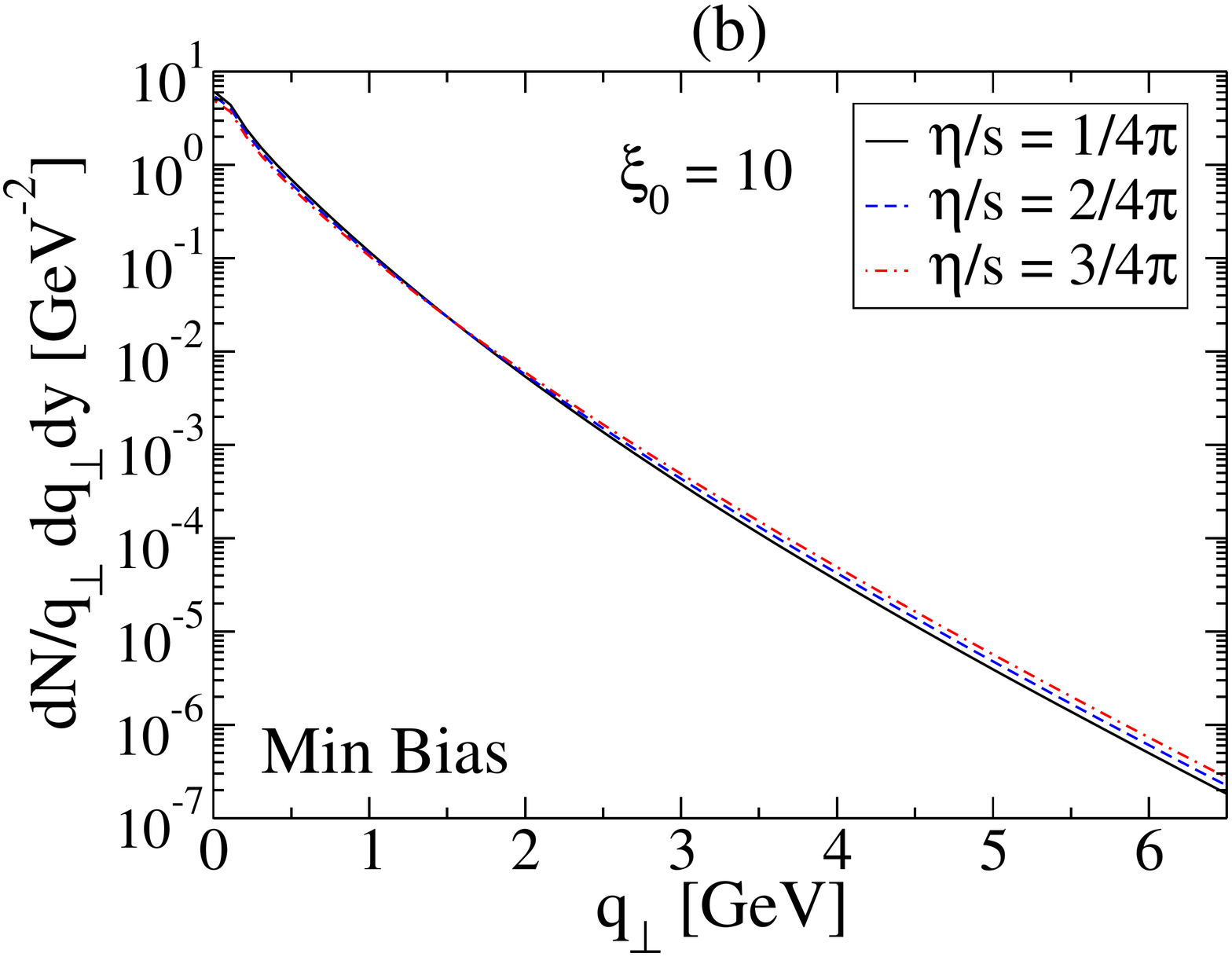}
\includegraphics[width=0.5\linewidth]{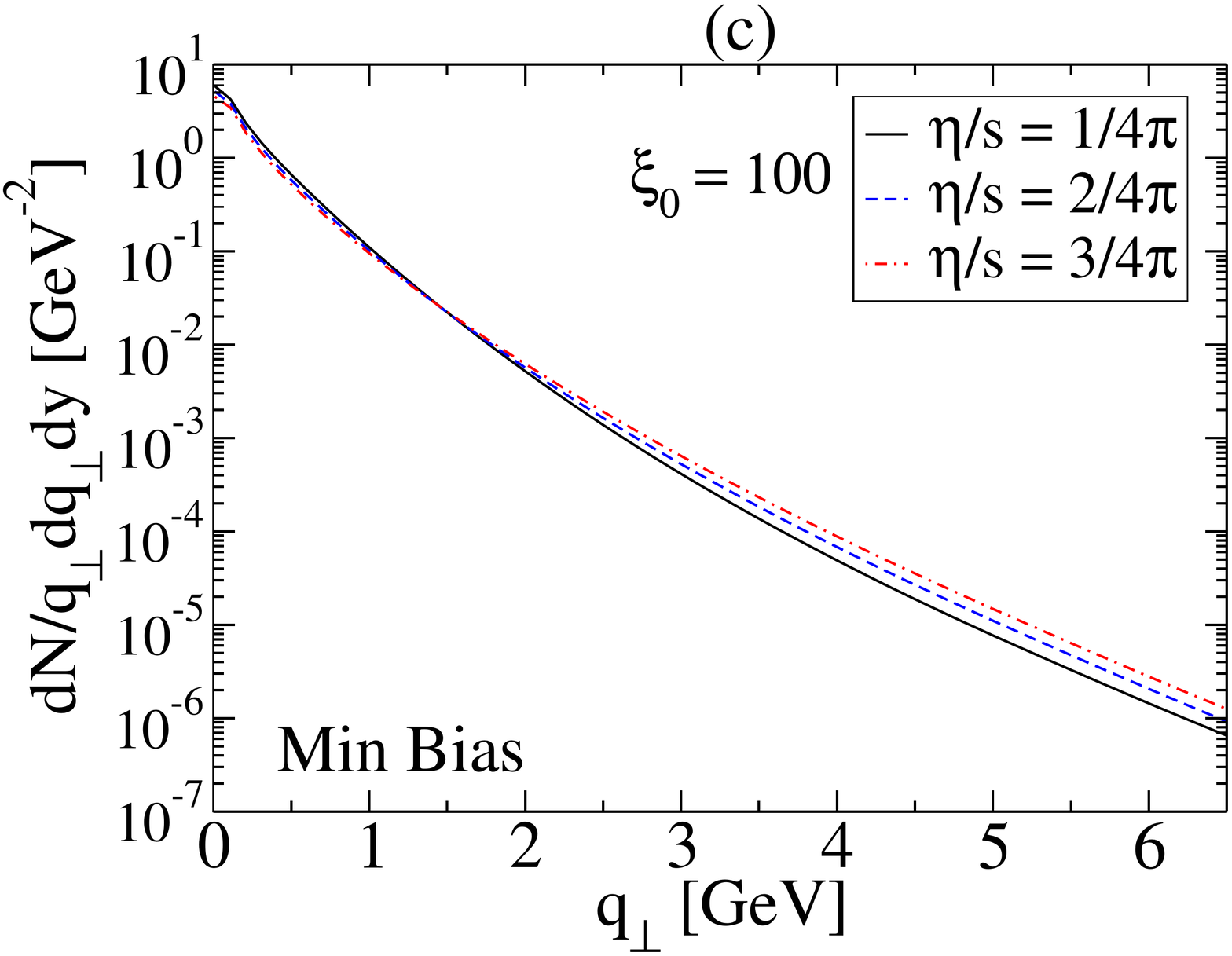}
\caption{Medium photon spectrum for three different values of the initial anisotropy: (a) $\xi_0 = 0$, (b) $\xi_0=10$, and (c) $\xi_0=100$.  In each panel, the lines correspond to three different values for the shear viscosity to entropy density ratio $4\pi \eta/s =$ 1, 2, and 3.}
\label{spectrafixmul}
\end{figure*}
%%%%%%%%%%%%%%%%%%%%%%%%%%%%%%%%%%%%%%%%%%%%%%%%%%%%%%%%%%%%%%%%%%%%%%%%%%%%%%%%%%%%%%%%%%%%%%%

%%%%%%%%%%%%%%%%%%%%%%%%%%%%%%%%%%%%%%%%%%%%%%%%%%%%%%%%%%%%%%%%%%%%%%%%%%%%%%%%%%%%%
\begin{figure*}[t]
\hspace{2mm}
\includegraphics[width=0.5\linewidth]{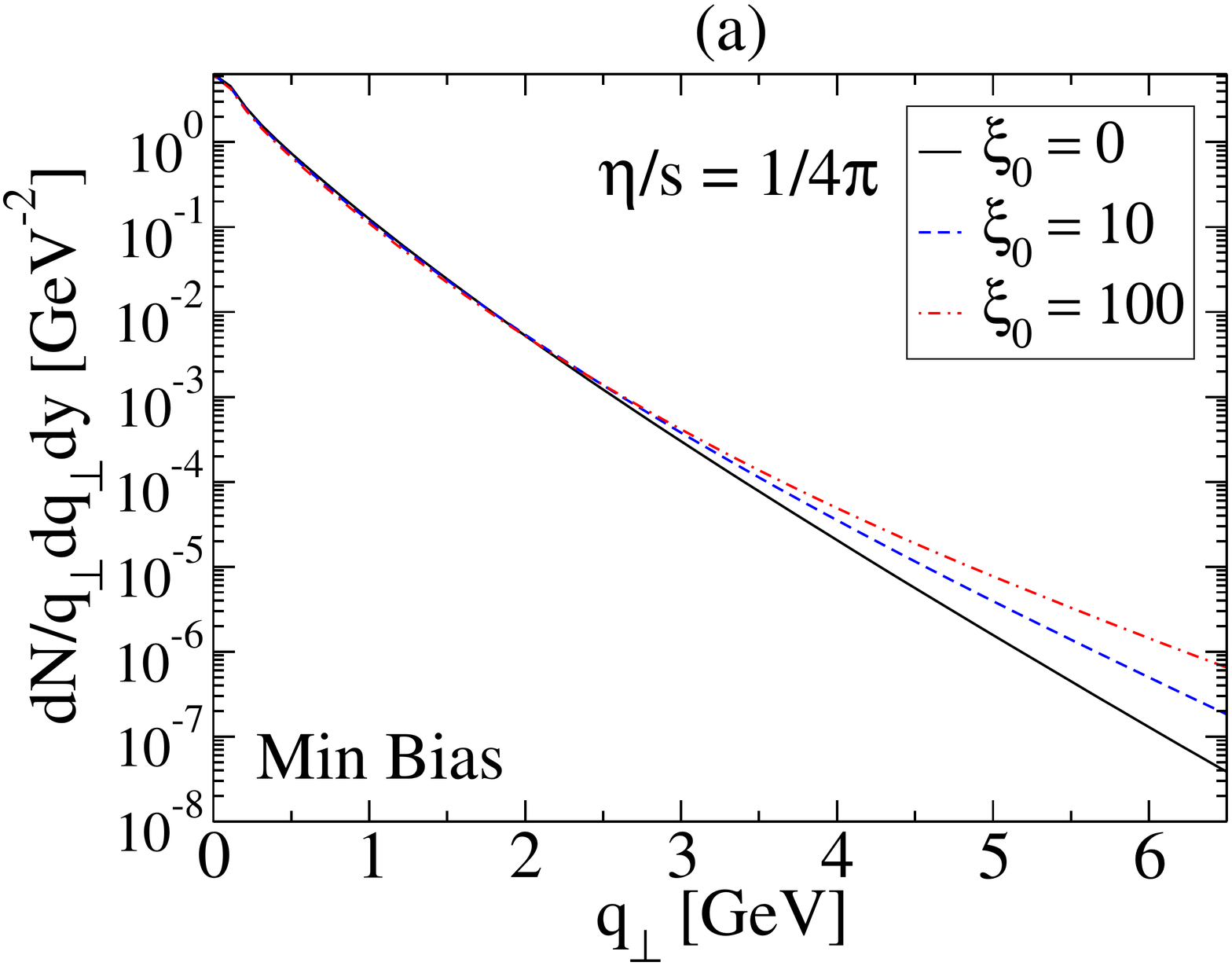}\hspace{-5mm}
\includegraphics[width=0.5\linewidth]{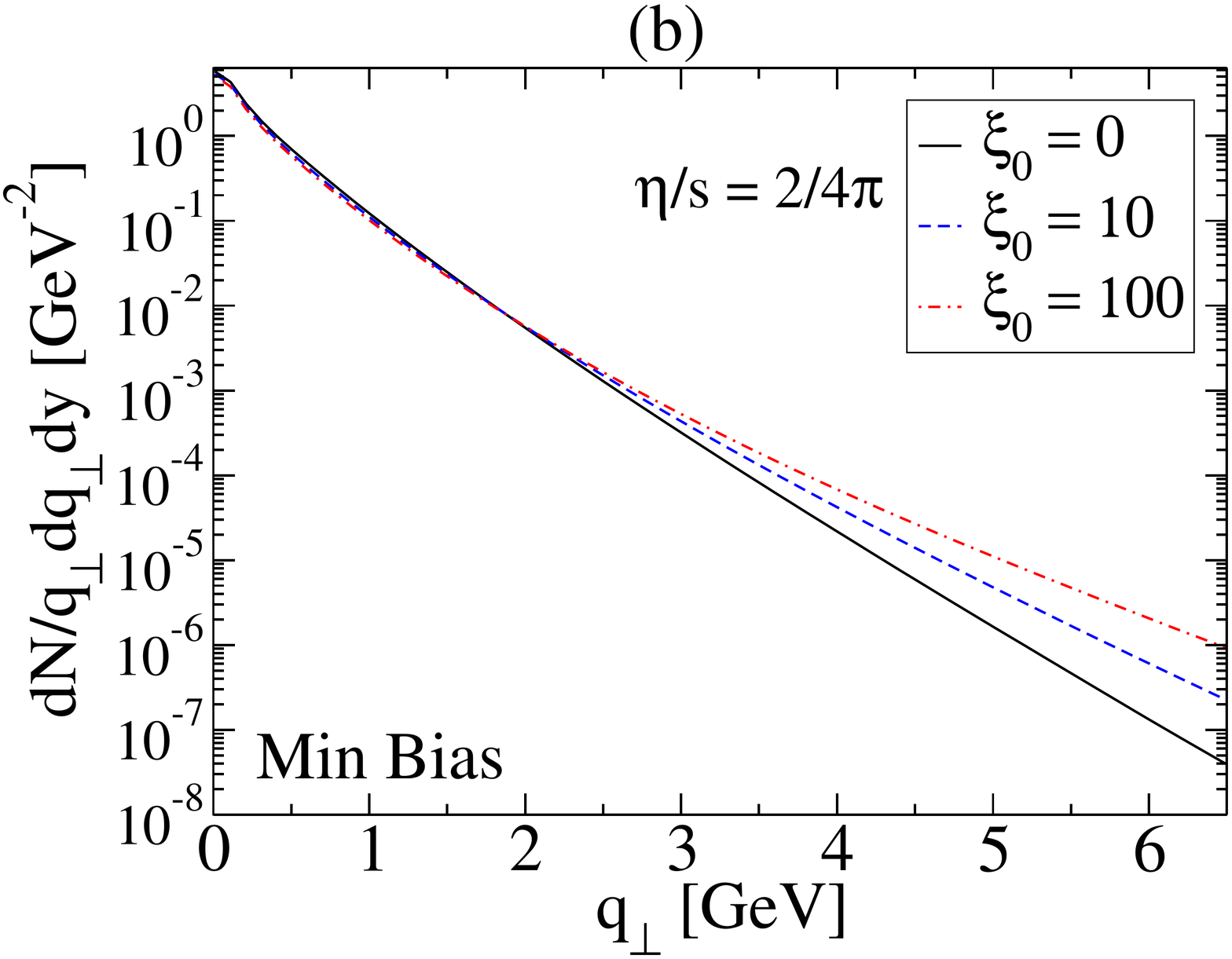}
\includegraphics[width=0.5\linewidth]{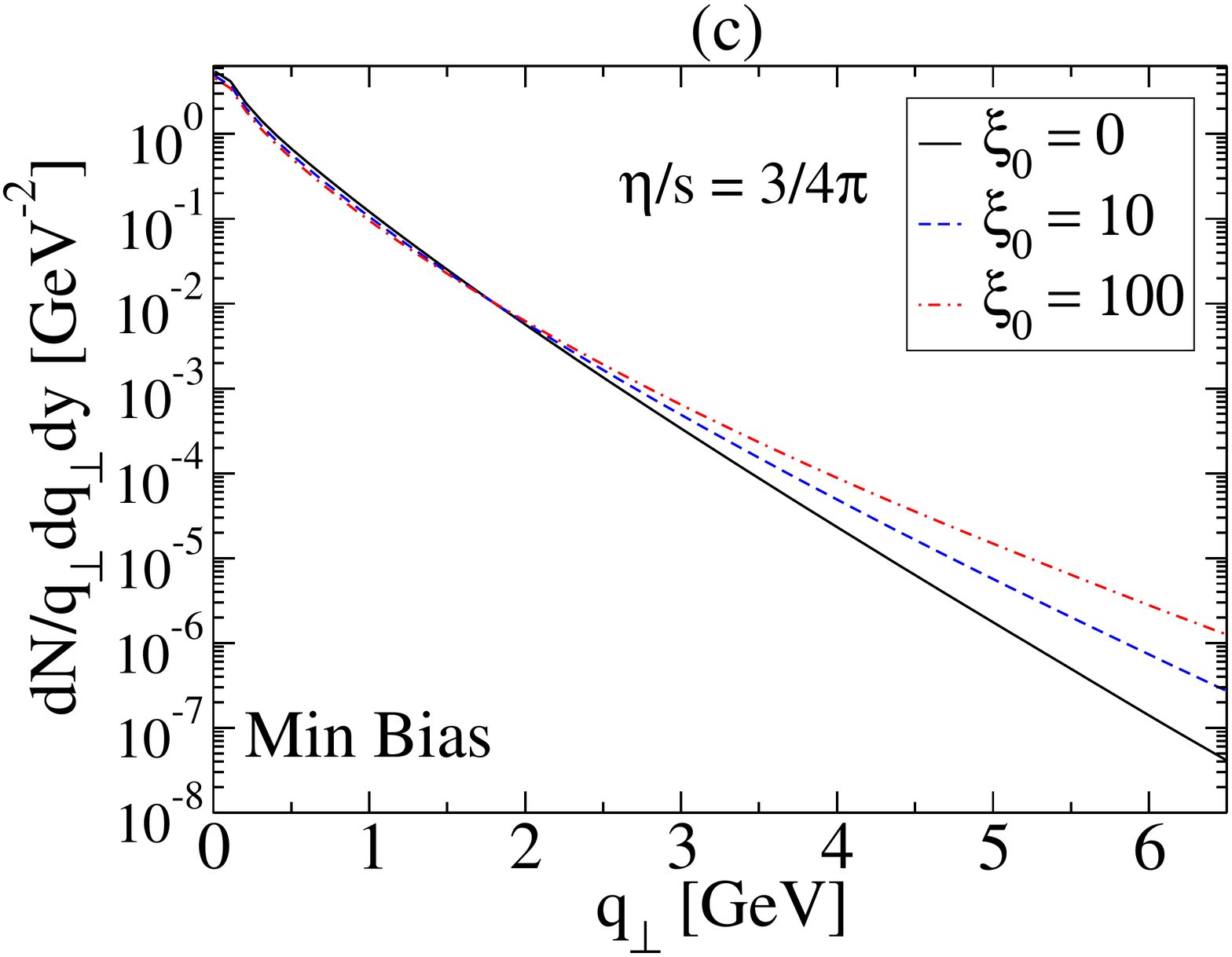}
\caption{Medium photon spectrum for three different values of viscosity (a) $4\pi\eta/s = 1$ , (b) $4\pi\eta/s = 2$ and (c) $4\pi\eta/s = 3$.  In each panel, the lines correspond to three different values for the 
initial anisotropy in the system $\xi_0 = 0, 10,$ and 100.}
\label{spectrafixmul2}
\end{figure*}
%%%%%%%%%%%%%%%%%%%%%%%%%%%%%%%%%%%%%%%%%%%%%%%%%%%%%%%%%%%%%%%%%%%%%%%%%%%%%%%%%%%%%%%%%%%%%%%

We begin with Fig.~\ref{spectrafixmul}, which shows the spectrum of medium photons obtained by integrating over the full (3+1)D evolution of the QGP.  The three different panels (a), (b), and (c) show the results obtained assuming initial anisotropies of $\xi_0 =$ 0, 10, and 100, respectively.  In each panel of Fig.~\ref{spectrafixmul}, the lines correspond to three different values of the shear viscosity to entropy density ratio $4\pi \eta/s =$ 1, 2, and 3.  As can be seen from panel (a), for $\xi_0=0$ the resulting spectrum is nearly the same for all three values of $\eta/s$, with only a very slight enhancement seen at large $q_\perp$.  The effect of varying $\eta/s$ is larger in panels (b) and (c) in which the QGP was assumed to have an initially oblate momentum-space anisotropy.  For $\xi_0=100$, there is approximately a factor 2.5 variation in the medium photon spectrum at $q_\perp$ = 6 GeV when varying $\eta/s$ between one and three times the lower bound.  For $q_\perp \lesssim$ 2 GeV, we see very little effect from varying $\eta/s$ for all values of $\xi_0$ considered.

In Fig.~\ref{spectrafixmul2} we present the same results in a slightly different manner.  In this case, in panels (a), (b), and (c) we fix the shear viscosity to entropy density ratio to be $4\pi\eta/s = 1$, $4\pi\eta/s = 2$, and $4\pi\eta/s = 3$, respectively.  In each of the panels of Fig.~\ref{spectrafixmul2} the lines correspond to three different values for the initial anisotropy in the system $\xi_0 = 0, 10,$ and 100.  As can be seen from Fig.~\ref{spectrafixmul2}, there is significant variation in the high-energy photon spectrum as one changes the initial anisotropy of the QGP.  For all values of $\eta/s$ considered, at \mbox{$q_\perp = 6$ GeV}, one finds that the QGP photon spectrum varies by approximately an order of magnitude when varying the initial anisotropy in the range shown.  We note additionally that for $q_\perp \lesssim$ 2 GeV, we see very little effect from varying the initial anisotropy.  Taken together, we see that the low-energy photon spectrum is not sensitive to either $\xi_0$ or $\eta/s$, if one keeps the final particle multiplicities fixed.

%%%%%%%%%%%%%%%%%%%%%%%%%%%%%%%%%%%%%%%%%%%%%%%%%%%%%%%%%%%%%%%%%%%%%%%%%%%%%%%%%%%%%
\begin{figure*}[t]
\hspace{2mm}
\includegraphics[width=0.5\linewidth]{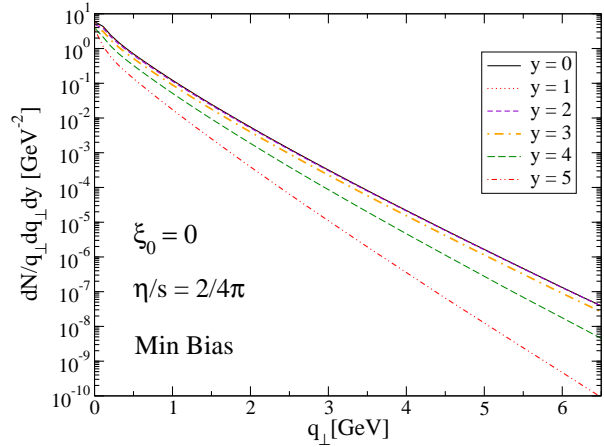}
\caption{Medium photon spectrum for six different values of the photon rapidity.  For this figure we assumed $\eta/s = 2/4\pi$ and $\xi_0=0$.}
\label{diffy_xi0}
\end{figure*}
%%%%%%%%%%%%%%%%%%%%%%%%%%%%%%%%%%%%%%%%%%%%%%%%%%%%%%%%%%%%%%%%%%%%%%%%%%%%%%%%%%%%%%%%%%%%%%%

In Fig.~\ref{diffy_xi0} we present the medium photon spectrum for six different values of the photon rapidity.  As this figure demonstrates for central rapidities, there is little dependence of photon production on the photon rapidity; however, for $y \gtrsim 3$ there is a significant dependence on the rapidity.  The fact that the central region is independent 
of the rapidity is consistent with the approximate boost-invariance of the quark-gluon plasma generated in heavy-ion collisions.  The dependence on rapidity at forward/backward rapidity is due to both the breaking of boost-invariance in the realistic 3+1d anisotropic hydrodynamics code and also the rapidity-dependence of the rate itself when the system is anisotropic (see Fig.~\ref{fig:rateFig}).  Note that, if the system was completely boost-invariant, then even with the rapidity dependence of the rate shown in Fig.~\ref{fig:rateFig}, one would find that photon production does not depend on rapidity.  We have verified this explicitly.

%%%%%%%%%%%%%%%%%%%%%%%%%%%%%%%%%%%%%%%%%%%%%%%%%%%%%%%%%%%%%%%%%%%%%%%%%%%%%%%%%%%%%
\begin{figure*}[t]
\hspace{2mm}
\includegraphics[width=0.5\linewidth]{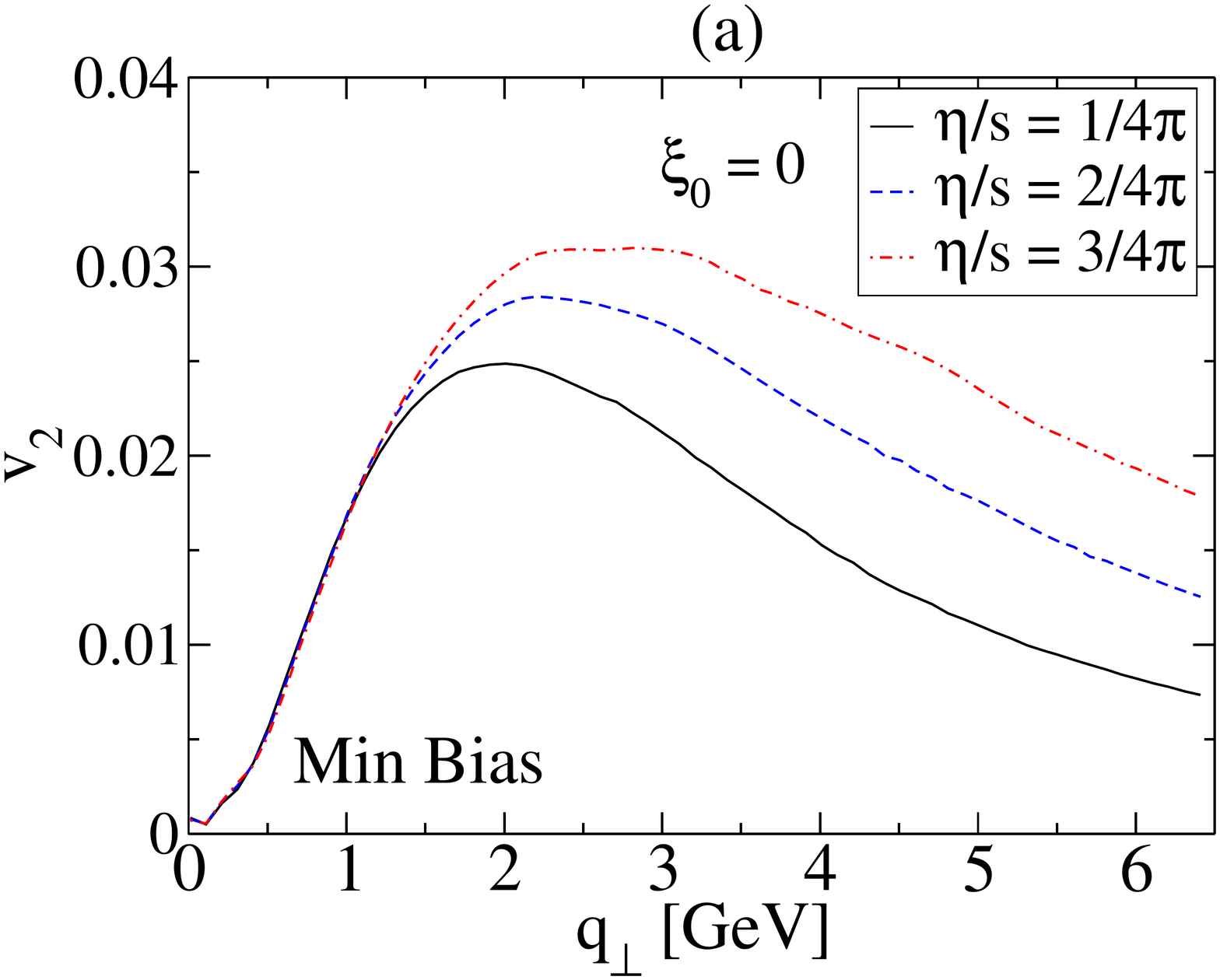}\hspace{-5mm}
\includegraphics[width=0.5\linewidth]{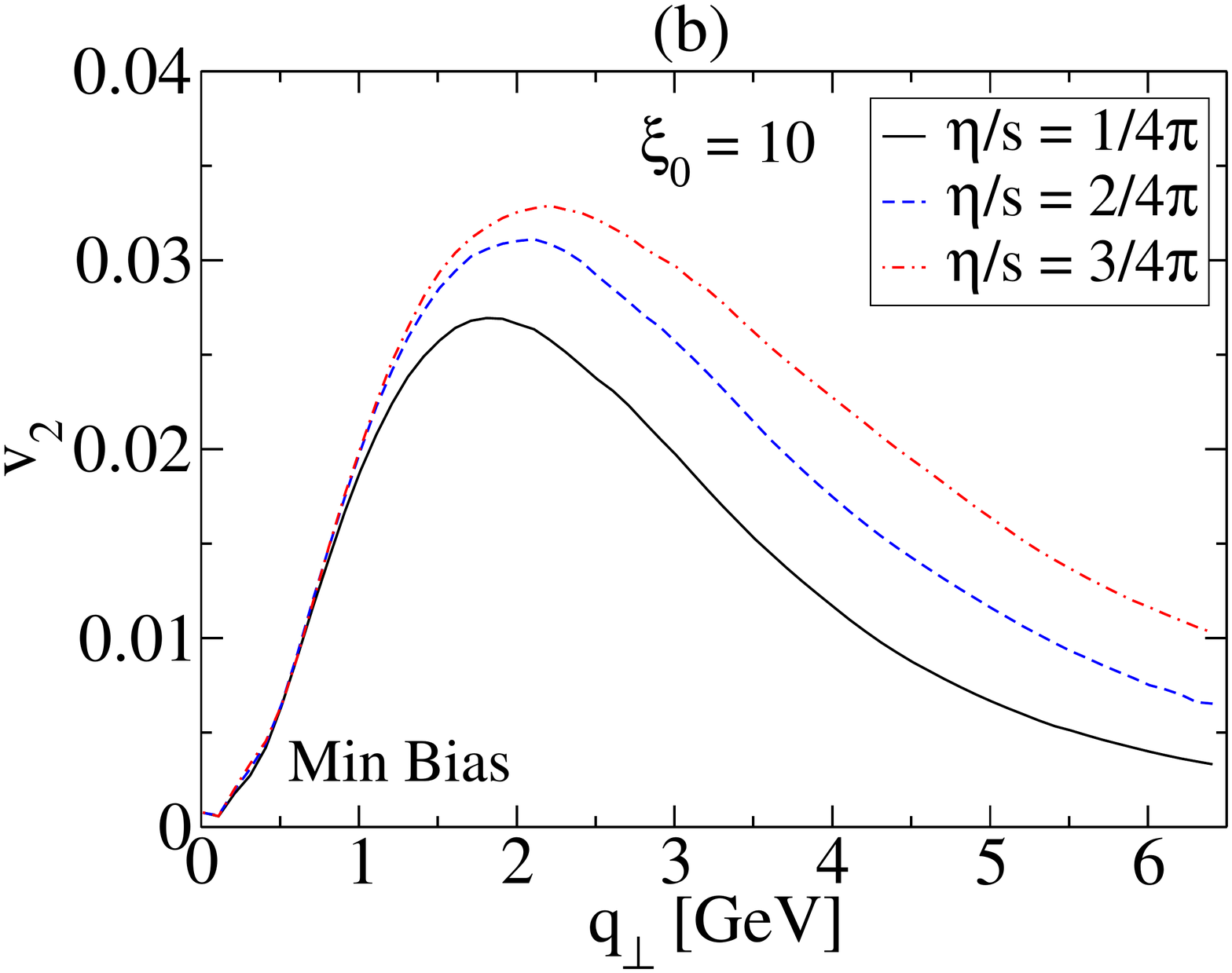}
\includegraphics[width=0.5\linewidth]{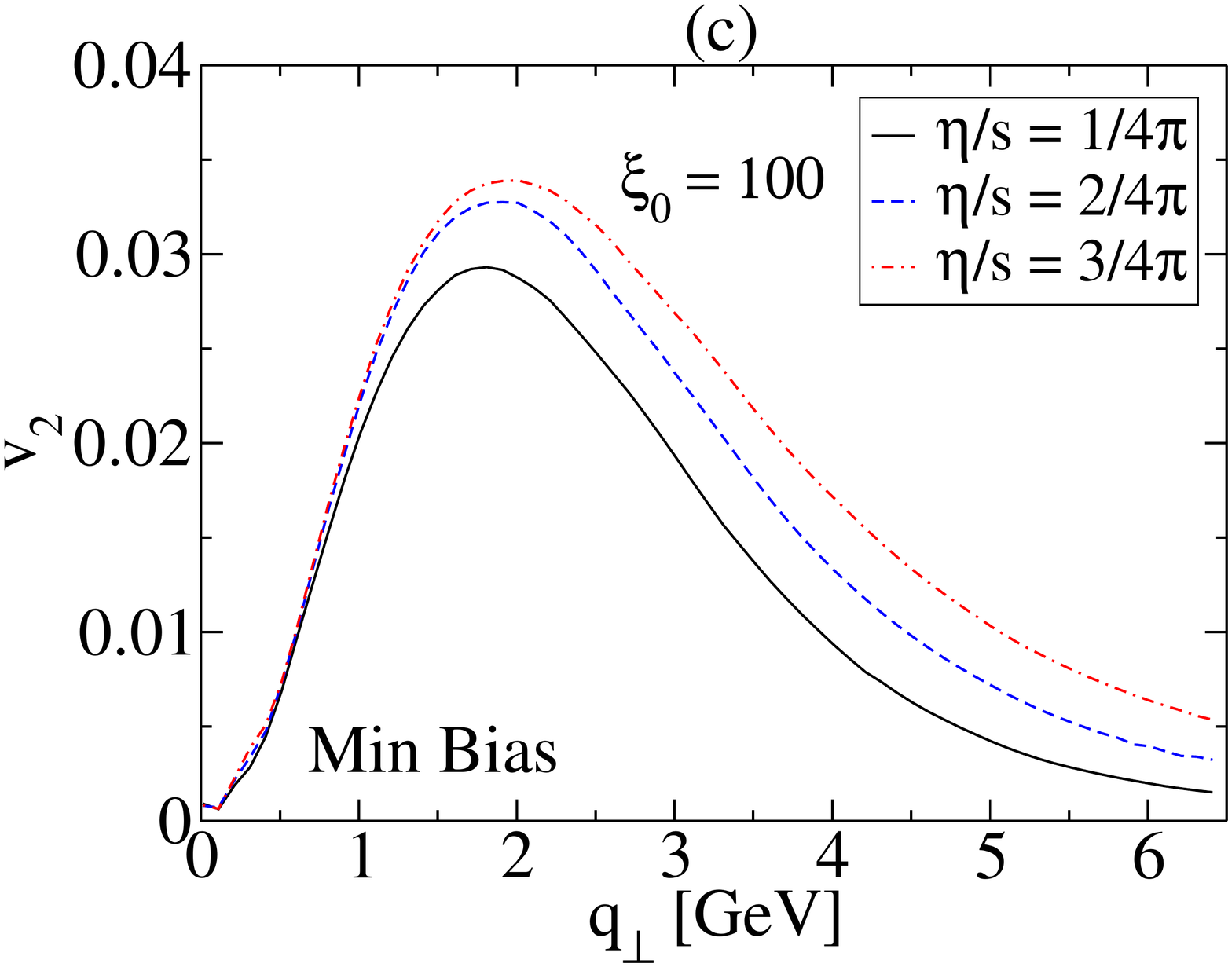}
\caption{Elliptic flow coefficient $v_2(q_\perp,y=0)$ for three different values of initial anisotropy 
parameter: (a) $\xi_0 = 0$, (b) $\xi_0 = 10$, and (c) $\xi_0 = 100$.  In each panel, the lines correspond to three different values for the shear viscosity to entropy density ratio $4\pi \eta/s =$ 1, 2, and 3.}
\label{V2_2}
\end{figure*}
%%%%%%%%%%%%%%%%%%%%%%%%%%%%%%%%%%%%%%%%%%%%%%%%%%%%%%%%%%%%%%%%%%%%%%%%%%%%%%%%%%%%%%%%%%%%%%%

We now turn to the photon elliptic flow which results from the breaking of rotational symmetry around the beam-axis due to the development of hydrodynamic flow.  To quantify the dependence of photon production on the momentum azimuthal angle, one makes a Fourier decomposition of the differential spectrum with respect to the momentum azimuthal angle
\begin{equation}
 \frac{dN}{q_\perp dq_\perp dy \, d\phi_q} = \frac{dN}{2\pi q_{\perp}dq_{\perp}dy}
 \Big[ 1+ 2 v_1 \cos{\phi_q} + 2 v_2 \cos(2\phi_q) + \ldots \Big] ,
\end{equation}
\label{v2}
where it is understood that the coefficients $v_n$ are functions of $q_\perp$ and $y$.\footnote{The coefficients are also implicit functions of the impact parameter, collision energy, colliding species, etc.}

From the above relation we can compute $v_2$ in the usual manner by extracting the second Fourier coefficient from the series via
\begin{equation}
v_2(q_{\perp},y) = \frac{ \int_0^{2\pi} d\phi_q \, \frac{dN}{q_\perp dq_\perp dy \, d\phi_q} \cos(2\phi_q) }{ \frac{dN}{q_\perp dq_\perp dy} } \, .
\label{V2_equation}
\end{equation}
This coefficient is referred to as the `elliptic flow coefficient', however, we emphasize that for photons a non-vanishing $v_2$ coefficient is not evidence of flow of the photons themselves, but instead the ``imprint'' of the transverse flow profile of the QGP itself.

%%%%%%%%%%%%%%%%%%%%%%%%%%%%%%%%%%%%%%%%%%%%%%%%%%%%%%%%%%%%%%%%%%%%%%%%%%%%%%%%%%%%
\begin{figure*}[t]
\centerline{\includegraphics[width=0.6\linewidth]{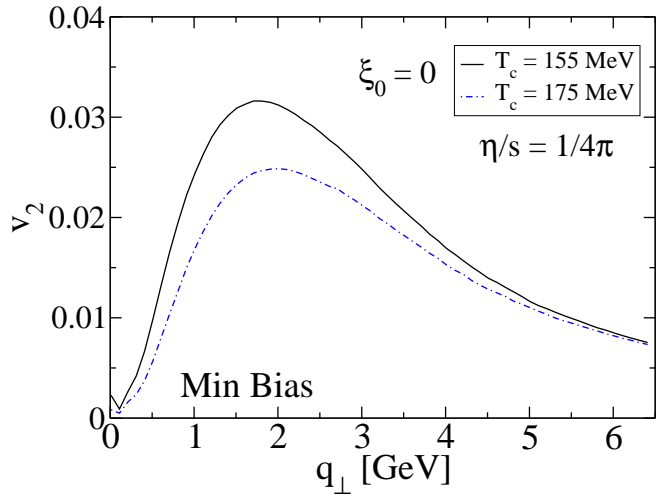}}
\caption{Variation of the photon $v_2$ with $T_c$.  For this figure we took $4\pi\eta/s=1$ and $\xi_0=0$.}
\label{V2diffTc}
\end{figure*}
%%%%%%%%%%%%%%%%%%%%%%%%%%%%%%%%%%%%%%%%%%%%%%%%%%%%%%%%%%%%%%%%%%%%%%%%%%%%%%%%%%%%%%%%%%%%%%%

In Fig.~\ref{V2_2}, we show our results for the photon $v_2$ coefficient in three panels (a), (b), and (c) which show the results obtained assuming initial anisotropies of $\xi_0 =$ 0, 10, and 100, respectively.  In each panel, we once again show the result obtained for three different values of the shear viscosity to entropy density ratio of $4\pi\eta/s = 1$, $4\pi\eta/s = 2$, and $4\pi\eta/s = 3$.  As Fig.~\ref{V2_2} demonstrates, we find that, regardless of the assumed initial momentum-space anisotropy, increasing the shear viscosity of the QGP results in an increase in photon $v_2$.  For the values of $\eta/s$ considered, we see at most a 300\% increase in the photon $v_2$ with the maximum effect occurring at high transverse momenta.  By comparing the three panels of Fig.~\ref{V2_2} we also see that, for fixed $\eta/s$, increasing the initial momentum-space anisotropy also results in an increase in the peak photon $v_2$, however, as $\xi_0$ increases there is a reduction in the photon $v_2$ at large transverse momentum.  In Fig.~\ref{V2diffTc}, we show the dependence of the photon $v_2$ on the assumed value for the transition temperature $T_c$.  In the figure we show the results obtained assuming $4\pi\eta/s=1$, $\xi_0=0$, and $T_c \in \{ 175, 155 \}$ MeV.  As we see from this figure, decreasing the critical temperature used when we integrate QGP emissions over the QGP four-volume results in an approximately 30\% increase in the peak value of the photon $v_2$.  In addition, we see that the peak in $v_2$ moves to lower transverse momentum as $T_c$ is decreased.

\begin{figure*}[t]
\centerline{
\hspace{3mm}
\includegraphics[width=0.52\linewidth]{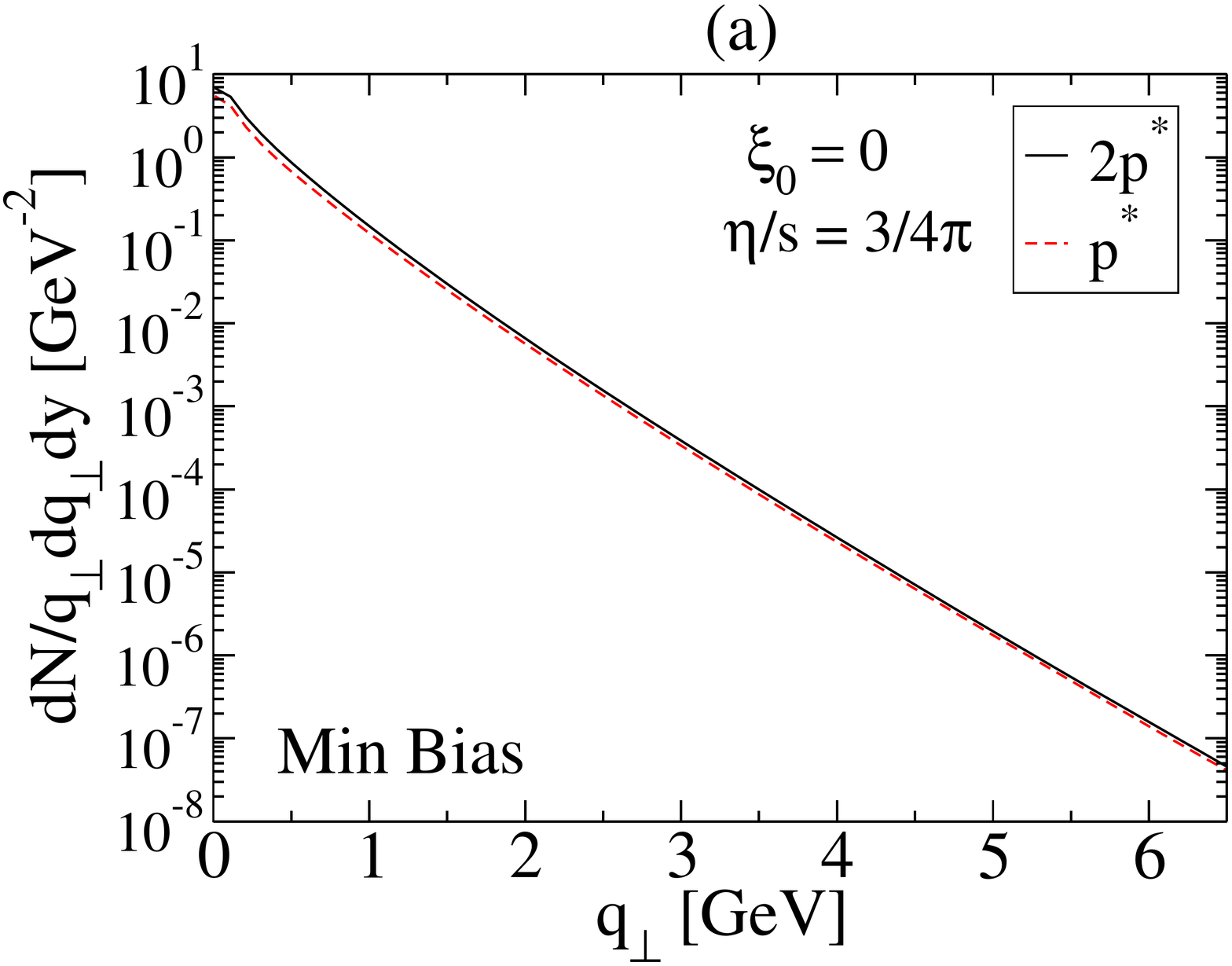}\hspace{-6mm}
\includegraphics[width=0.52\linewidth]{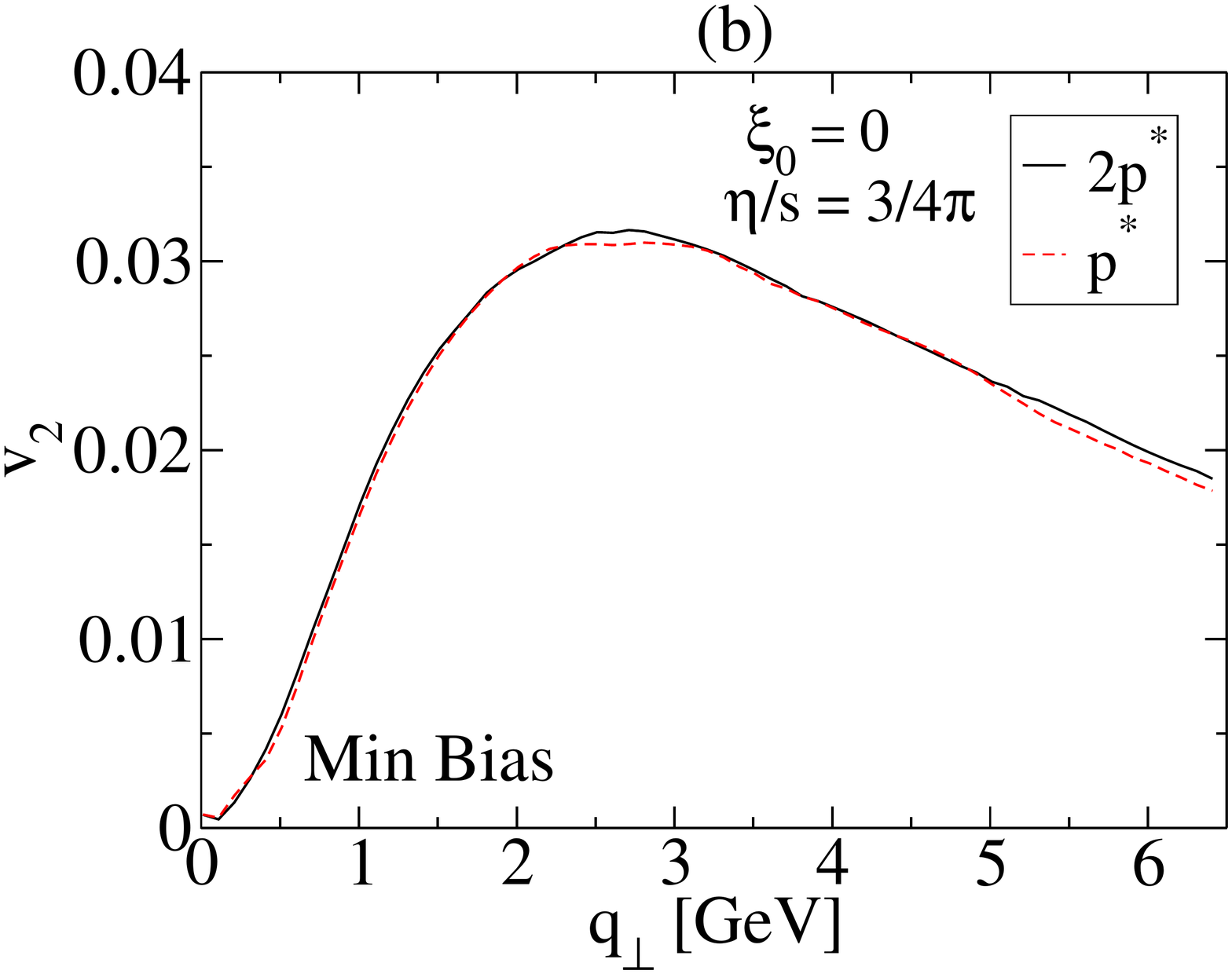}
}
\caption{Dependence of (a) the medium photon spectrum and (b) the elliptic flow coefficient on transverse momentum for two different values of separation scale:  $p^*$ and $2p^*$.  In both panels, we assumed that the $\xi_0=0$ and $4\pi\eta/s=3$.}
\label{V2pstar}
\end{figure*}

Finally, as a cross check of the results shown thus far, in Fig.~\ref{V2pstar} we show (a) the photon spectrum and (b) the photon elliptic flow as a function of the transverse momentum comparing the result obtained if we use the standard PMS criteria to set the soft/hard separation scale $p^*$ (red dashed line) or instead take the separation scale to be two times the PMS value (black line).  As Fig.~\ref{V2pstar}(a) demonstrates, using $2p^*$ results in a higher photon yield by approximately 30-50\%.  However, as Fig.~\ref{V2pstar}(b) demonstrates, the additional production obtained when using $2p^*$ cancels in the ratio that determines $v_2$ and, as a result, the photon $v_2$ is independent of the choice of the separation scale within numerical uncertainties.\footnote{The small differences between the two curves in Fig.~\ref{V2pstar} can be attributed to statistical errors inherent in the Monte-Carlo integration method used to evaluate the final spectrum.}  Finally, we mention that, although we only show the case $\xi_0=0$ and  $4\pi\eta/s=3$ in Fig.~\ref{V2pstar}, for all $\xi_0$ and $\eta/s$ considered here, we find that $v_2$ does not depend on the separation scale.

%%%%%%%%%%%%%%%%%%%%%%%%%%%%%%%%%%%%%%%%%%%%%%%%%%%%%%%%%%%%%%%%%%%%%%%%%%%%%%%%%%%%
%%%%%%%%%%%%%%%%%%%%%%%%%%%%%%%%%%%%%%%%%%%%%%%%%%%%%%%%%%%%%%%%%%%%%%%%%%%%%%%%%%%%%%%%%%%%%%%

%%%%%%%%%%%%%%%%%%%%%%%%%%%%%%%%%%%%%%%%%%%%%%%%%%%%%%%%%%%%%%%%%%%%%%%%%%%%%%%%%%%%%%%%%%%%%%%
\section{Discussion, conclusions, and outlook}
\label{sec:conclusions}
%%%%%%%%%%%%%%%%%%%%%%%%%%%%%%%%%%%%%%%%%%%%%%%%%%%%%%%%%%%%%%%%%%%%%%%%%%%%%%%%%%%%%%%%%%%%%%%

In this paper we computed the photon spectrum and elliptic flow coefficient associated with photons emerging from the QGP as a function of transverse momentum.  For the rate, we included the two leading-order processes necessary: Compton scattering and quark-antiquark annihilation.  In order to properly deal with the IR divergences encountered in the rate calculation, we computed the soft and hard contributions separately.  In the soft sector, the UV divergence was regulated by introducing a UV cutoff $p^*$ and, in the hard sector, the IR divergence was regulated by introducing an IR cutoff $p^*$.  As shown analytically by Braaten and Yuan \cite{Braaten:1991dd}, in the weak-coupling limit, the sum of the hard and soft contributions is finite due to a cancellation between the UV and IR divergences and the result does not depend on the separation scale $p^*$.  However, for realistic couplings, one must evaluate the necessary integrals numerically.  In this case, there is a residual dependence on the choice of the separation scale.  To fix it, we used a PMS criteria to set $p^*$ to be the scale at which the derivative of the rate vanishes, thereby minimizing the dependence of the rate on the separation scale.  

After determining the rates, we then integrated them over the space-time volume of the QGP using (3+1)D anisotropic hydrodynamics to provide the space and time dependence of the anisotropy $\xi(x)$, the transverse momentum scale $\Lambda(x)$, and the flow velocity $u^\mu(x)$.  In the aHydro framework, the one-particle distribution function is guaranteed to be greater than or equal to zero unlike in the standard viscous hydrodynamic framework where there can be regions in phase space where the one-particle distribution function is negative.  Our final results indicate that, if one holds the final particle multiplicity fixed, there is only a weak dependence of the photon spectrum on the assumed value of $\eta/s$.  However, we found that, for fixed $\eta/s$, varying the initial momentum-space anisotropy $\xi_0$ resulted in significant variations of the high-transverse-momentum photon yields.  We found that, at $q_\perp = 6$ GeV, there was approximately an order of magnitude variation in the photon yield when varying $0 \leq \xi_0 \leq 100$.  This offers some hope to constrain the degree of QGP momentum-space anisotropy by fitting thermal plus prompt photon production at high energies to experimental data.  We note that this is similar to the conclusion reached recently for dilepton production, where an enhancement of the high-energy dilepton production was observed when the initial condition was initially oblate in momentum space~\cite{Ryblewski:2015hea}.

In addition to presenting results for the spectrum, in this paper we also calculated the elliptic flow coefficient $v_2$ associated with the azimuthal variation of the photon yields.  Our results indicate that, for $4\pi\eta/s=1$, the maximal photon $v_2$ coming from the QGP phase at LHC energies is approximately 2-3\%.  We find that increasing $\eta/s$ or $\xi_0$ results in an increase in peak photon $v_2$.  Our finding that $v_2$ increases as $\eta/s$ is increased seems to be in conflict with some earlier papers, e.g. \cite{Dion:2011pp,Shen:2013vja,Shen:2013cca,Shen:2014nfa}, which found that incorporating viscous corrections resulted in a decrease in photon $v_2$.  One possible explanation for the different trends in the photon $v_2$ seen using aHydro versus second-order viscous hydrodynamics is that herein we employed the leading-order spheroidal form for the LRF distribution function, allowing for only one dissipative correction, quantified by $\xi$, which maps to the longitudinal-transverse pressure anisotropy.  In second-order viscous hydrodynamics, the description of the shear tensor is more complete, i.e. transverse pressure anisotropies and off-diagonal terms are present, resulting in a total of five independent degrees of freedom.  For quantitative assessment of photon $v_2$, the additional viscous corrections could be important.  
Finally, we also note that our study is quite limited since we did not include fluctuating initial conditions or hadronic sources of photons.  These are both major shortcomings of this work.  Our intention herein was to study the systematics in a simply context in which the anisotropic screening and one-particle distribution functions were both taken into account self-consistently.  We plan to include multiple anisotropy parameters, fluctuating initial conditions, and hadronic emissions in forthcoming papers.
Importantly, however, we point out that the dependence of the photon spectrum on the initial anisotropy $\xi_0$ found here is primarily sensitive to early-time longitudinal-transverse pressure anisotropies with the other viscous corrections being subleading.  Therefore, we are confident that this effect is generic and reasonably well-described using leading-order aHydro.

We mention in closing that, during the analysis, we showed that there is an approximately 30\% increase in the QGP photon spectrum when varying the hard/soft separation scale by a factor of two, however, we found that the photon $v_2$ did not depend on the choice of the separation scale.  In the future, in addition to improving upon the aHydro assumptions, we will combine the results obtained here with estimates for prompt photon production in order to extract constraints on early-time momentum-space anisotropies in the QGP.

\acknowledgments{We thank G.~Denicol, C.~Shen, U.~Heinz, and J.~Paquet for useful conversations.  L.~Bhattacharya was supported by a M.~Hildred Blewett Fellowship from the American Physical Society.  M.~Strickland was supported by the U.S.~Department of Energy under Award No.~DE-SC0013470.  R. Ryblewski was supported by the Polish National Science Center Grant No.~DEC-2012/07/D/ST2/02125.}

\bibliography{photon_3d}

\end{document}